\documentclass[12pt]{article}
\usepackage{amssymb,amsmath,latexsym}
\title{Geometry of Weyl theory
\\ for Jacobi matrices with matrix entries}

\author{Hermann Schulz-Baldes
\\
\\
{\small Department Mathematik, Universit\"at Erlangen-N\"urnberg, Germany}
}

\date{ }

\newtheorem{theo}{Theorem}
\newtheorem{defini}{Definition}
\newtheorem{proposi}{Proposition}
\newtheorem{lemma}{Lemma}
\newtheorem{coro}{Corollary}

\newcommand{\CC}{{\mathbb C}}
\newcommand{\NN}{{\mathbb N}}
\newcommand{\RR}{{\mathbb R}}

\newcommand{\LM}{{\mathbb L}}

\newcommand{\DM}{{\mathbb D}}
\newcommand{\GM}{{\mathbb G}}
\newcommand{\UM}{{\mathbb U}}
\newcommand{\WM}{{\mathbb W}}

\newcommand{\Pp}{{\cal P}}

\newcommand{\HH}{{\bf H}}
\newcommand{\UU}{{\bf U}}

\newcommand{\Dd}{{\cal D}}

\newcommand{\Gg}{{\cal G}}
\newcommand{\Ww}{{\cal W}}
\newcommand{\Qq}{{\cal Q}}

\newcommand{\Tt}{{\cal T}}
\newcommand{\Rr}{{\cal R}}

\newcommand{\Jj}{{\cal J}}

\newcommand{\Hh}{{\cal H}}

\newcommand{\one}{{\bf 1}}
\newcommand{\nul}{{\bf 0}}
\newcommand{\inv}{{\mbox{\rm\tiny inv}}}

\newcommand{\mx}{{\mbox{\rm\tiny max}}}

\begin{document}

\maketitle

\begin{abstract}
A Jacobi matrix with matrix entries is a self-adjoint block tridiagonal matrix
with invertible blocks on the off-diagonals. The Weyl surface describing the
dependence of Green's matrix on the boundary conditions is interpreted as the
set of maximally isotropic subspace of a quadratic from given by the Wronskian.
Analysis of the possibly degenerate limit quadratic form leads to the limit
point/limit surface theory of maximal symmetric extensions for semi-infinite
Jacobi matrices with matrix entries with arbitrary deficiency indices. The
resolvent of the extensions is explicitly calculated.
\end{abstract}

\vspace{.5cm}

\section{Introduction}

Based on the work of Krein \cite{Kre}, Berezanskii's monograph \cite{Ber}
describes a Weyl-Titchmarch-type theory of self-adjoint extensions for
semi-infinite Jacobi matrices with matrix entries. However, the nice geometric
picture of the Weyl circle was fully extended to the matrix case only by
Fukushima \cite{Fuk}, and for Hamiltonian systems by Schneider, Niessen, Hinton
and Shaw (in a series of papers, see \cite{Sch,Nie0,HS,HSch} and references
therein; the recent paper by Clark and Gesztesy \cite{CG} contains many further
references). Other earlier works on similar operators defined by linear
differential systems are \cite{Nai,Wei}. Here the theory is developed in detail
for Jacobi matrices with matrix entries. Following the reasoning of Simon
\cite{Sim}, the results below are relevant for the matrix moment problem (see
\cite{DL} for a review).

\vspace{.2cm}

The basic fact is that for any complex energy $z\in\CC/\RR$ the Green matrix at
finite volume $N$ considered as a function of the boundary conditions varies in
a compact set of matrices which is diffeomorphic the to the unitary group (see
Theorem~\ref{theo-Weyl} below). This set is called the {\it Weyl surface}
$\partial_L\WM^z_N$ and it can be identified with the maximally isotropic
subspaces of a quadratic form $\Qq^z_N$ defined by the Wronskian of the
transfer matrices (Section~\ref{sec-isotropic}). It is also helpful to think of
the Weyl surface as the maximal boundary of the so-called Weyl disc (similar as
the unitary group is the maximal boundary of Cartan's first classical domain).
The Weyl disc and its surface always shrink and are nested as the volume grows
(Theorem~\ref{theo-Weyl}). Hence there exists a Weyl limit surface
$\partial\WM^z$ as $N\to\infty$ (Section~\ref{sec-semiinfinite}). There exists
also a limit quadratic form $\Qq^z$ and its maximally isotropic subspaces can
again be identified with the Weyl limit surface
(Proposition~\ref{prop-Weylsurfacedescription}). In the scalar case there is a
simple dichotomy associating limit point behavior to an essentially
self-adjoint operator and limit circle behavior to an operator with deficiency
indices $(1,1)$. In the matrix case the situation is more rich. For example, it
is possible that the deficiency indices are not equal even though one of the
Weyl limit discs consists of only one point. For real Jacobi matrices the
deficiency indices are always equal though.

\vspace{.2cm}

All possible symmetric extensions can be constructed by standard von Neumann
theory and the Weyl limit surface $\partial_\mx\WM^\zeta$ associated to a fixed
complex energy $\zeta$ parametrizes all of them (Theorem~\ref{theo-isospaces}).
It can then be shown that this also fixes a unique maximally isotropic subspace
of $\Qq^z$ for any other $z$ in the upper half-plane (when the deficiency index
in the upper half-plane is smaller than or equal to that in the lower
half-plane). This subspace therefore fixes a unique point also on the Weyl
surface $\partial_\mx\WM^z$. This point is precisely the Green matrix at $z$.
The proof heavily uses the limit Wronskian in order to characterize the
deficiency spaces (Theorems~\ref{theo-domainchar} and \ref{theo-isochar}). This
geometric argument allows to calculate the resolvent of the extensions
explicitly (details are given after Theorem~\ref{theo-greencalc}). The obtained
formula for the resolvent differs from the one by Nevalina rederived in
\cite{Sim}.

\vspace{.2cm}

As already indicated above, there exists a large literature on the topics of
this paper, and the author is sure to only know a fraction of it. This paper
does provide a self-contained description of the generalization of
Weyl-Titchmarch theory to Jacobi matrices with matrix entries allowing all
possible values of the deficiency indices. It streamlines the proofs and
stresses the structural and geometric aspects of the theory. On the other hand,
analyticity of the various objects in real and imaginary part of the energy is
not tracked in detail.

\vspace{.2cm}

This work complements \cite{SB} which discussed Sturm-Liouville oscillation
theory for Jacobi matrices with matrix entries (also called discrete 
Sturm-Liouville operators \cite{AN}), so the basic notations and
setup are chosen accordingly. At some points use is being made of the matrix
M\"obius transformation on which there is an abundant literature (see the
references in \cite{DPS}), but the main facts relevant for the present purposes
are resembled in an appendix and all their short proofs are given in \cite{SB}.

\vspace{.2cm}

\noindent {\bf Notations:} The matrix entries of the Jacobi matrices are of
size $L\in\NN$. Matrices of size $L\times L$ are denoted by roman letters,
those of size $2L\times 2L$ by calligraphic ones. The upper half-plane $\UM_L$
is the set of complex $L\times L$ matrices satisfying $\imath(Z^*-Z)>\nul$. Its
closure $\overline{\UM_L}$ is given by matrices satisfying
$\imath(Z^*-Z)\geq\nul$. The boundaries is a stratified space
$\partial\UM_L=\cup_{l=1}^L\partial_l\UM_L$, where $\partial_l\UM_L$ contains
those matrices in $\overline{\UM_L}$ for which the kernel of $Z^*-Z$ is
$l$-dimensional.

\vspace{.3cm}

\noindent {\bf Acknowledgment:} This work was supported by the DFG.

\vspace{.3cm}

\section{Solutions of the finite system} \label{sec-finite}

Fix two integers $L,N\in \NN$ and let $(T_n)_{n=2,\ldots,N}$ and
$(V_n)_{n=1,\ldots,N}$ be sequences of respectively invertible and self-adjoint
$L\times L$ matrices with complex entries. Furthermore let the left and right
boundary conditions $\hat{Z}$ and $Z$ be also self-adjoint $L\times L$
matrices. Then the associated Jacobi matrix with matrix entries
$H^N_{\hat{Z},Z}$ is by definition the self-adjoint operator acting on states
$\phi=(\phi_n)_{n=1,\ldots,N}\in \ell^2(1,\ldots,N)\otimes \CC^L$ by
\begin{equation}
\label{eq-jacobi} (H^N_{\hat{Z},Z}\,\phi)_n \;=\;
T_{n+1}\phi_{n+1}\,+\,V_n\phi_n \,+\,T_{n}^*\phi_{n-1} \;, \qquad
n=1,\ldots,N\;,
\end{equation}
where $T_1=T_{N+1}={\bf 1}$, together with the boundary conditions
\begin{equation}
\label{eq-boundary} \phi_0\;=\;\hat{Z}\,\phi_1\;, \qquad
\phi_{N+1}\;=\;-\,Z\,\phi_N\;.
\end{equation}
If $\hat{Z}={\bf 0}$ and ${Z}={\bf 0}$, one speaks of Dirichlet boundary
conditions at the left an right boundary respectively. It will be useful to
allow also non-selfadjoint boundary conditions $\hat{Z},Z\in\overline{\UM_L}$
hence giving rise to a possibly non-selfadjoint operator $H^N_{\hat{Z},Z}$. One
can rewrite $H^N_{\hat{Z},Z}$ as an $NL\times NL$ matrix with $L\times L$ block
entries:
\begin{equation}
\label{eq-matrix} H^N_{\hat{Z},Z} \;=\; \left(
\begin{array}{ccccccc}
V_1-\hat{Z}       & T_2  &        &        &         &        \\
T_2^*      & V_2    &  T_3  &        &         &        \\
            & T_3^* & V_3    & \ddots &         &        \\
            &        & \ddots & \ddots & \ddots  &        \\
            &        &        & \ddots & V_{N-1} & T_N   \\
            &        &        &        & T_N^*  & V_N -Z
\end{array}
\right)
\;.
\end{equation}
At times, our interest will only be in the dependence of the right boundary
condition $Z$, and then the index $\hat{Z}$ will be suppressed.

\vspace{.2cm}

As for a one-dimensional Jacobi matrix, it is useful to rewrite the eigenvalue
equation
\begin{equation}
\label{eq-Schroedinger} (H^N_{Z,\hat{Z}}\,\phi)_n \;=\; z\,\phi_n\;, \qquad
n=1,\ldots,N\;,
\end{equation}
for a complex energy $z\in \CC$
in terms of the $2L\times 2L$ transfer matrices $\Tt_n^z$ defined by
\begin{equation}
\label{eq-transfer}
\Tt_n^z
\;=\;
\left(
\begin{array}{cc}
(z\,{\bf 1}\,-\,V_n)\,T_n^{-1} & - T_n^* \\
T_n^{-1} & {\bf 0}
\end{array}
\right)
\;,
\qquad
n=1,\ldots,N
\;,
\end{equation}
namely
\begin{equation}
\label{eq-transfersol}
\left(
\begin{array}{c}
T_{n+1}\phi_{n+1} \\
\phi_n
\end{array}
\right)
\;=\;
\Tt^z_n\,
\left(
\begin{array}{c}
T_{n}\phi_{n} \\
\phi_{n-1}
\end{array}
\right)
\;,
\qquad
n=1,\ldots,N
\;.
\end{equation}
This gives a solution of the eigenvalue equation \eqref{eq-Schroedinger} which,
however, does not necessarily satisfy the boundary condition
\eqref{eq-boundary}. Now $z\in\CC$ is an eigenvalue of $H^N_{\hat{Z},Z}$ if and
only if there is a solution of \eqref{eq-Schroedinger}, that is produced by
\eqref{eq-transfersol}, which satisfies \eqref{eq-boundary}. As is
well-established, one can understand \eqref{eq-boundary} as requirement on the
solution at sites $0,1$ and $N,N+1$ respectively to lie in $L$-dimensional
planes in $\CC^{2L}$. The corresponding two planes are described by the two
$2L\times L$ matrices (one thinks of the $L$  columns as spanning the plane)
\begin{equation}
\label{eq-boundaryplanes} \hat{\Phi}_{\hat{Z}} \;=\; \left(
\begin{array}{c}
{\bf 1} \\
-\,\hat{Z}
\end{array}
\right)\;, \qquad \Phi_Z \;=\; \left(
\begin{array}{c}
-\,Z \\
{\bf 1}
\end{array}
\right)\;.
\end{equation}
Then the boundary conditions (\ref{eq-boundary}) can be rewritten as
\begin{equation}
\label{eq-boundary2}
\left(
\begin{array}{c}
T_1\phi_{1} \\
\phi_0
\end{array}
\right) \;\in\; \hat{\Phi}_{\hat{Z}} \,\CC^L \;, \qquad \left(
\begin{array}{c}
T_{N+1}\phi_{N+1} \\
\phi_N
\end{array}
\right) \;\in\; \Phi_Z \,\CC^L \;.
\end{equation}
One way to attack the eigenvalue problem is to consider the $L$-dimensional
plane $\hat{\Phi}_{\hat{Z}}$ as the initial condition for an evolution of
$L$-dimensional planes under the application of the transfer matrices:
\begin{equation}
\label{eq-Lagdyn} \Phi_n^z \;=\; \Tt_n^z\,\Phi_{n-1}^z \;, \qquad \Phi_0^z
\;=\; \hat{\Phi}_{\hat{Z}} \;.
\end{equation}
Because the transfer matrices are invertible, this produces an $L$-dimensional
set of solutions of \eqref{eq-transfersol}. With the correspondence
\begin{equation}
\label{eq-solcorr} \Phi^{z}_n\;=\; \left(
\begin{array}{c}
T_{n+1}\, \phi^{z}_{n+1}\\
\phi^{z}_n
\end{array}
\right)\;,
\end{equation}
this also gives a matricial solution $\phi_n^z$ of \eqref{eq-Schroedinger}. Due
to the initial condition in \eqref{eq-Lagdyn} the left boundary condition at
sites $0,1$ is automatically satisfied. The dimension of the intersection of
the plane $\Phi_N^z$ with the plane $\Phi_Z$ gives the number of linearly
independent solutions of \eqref{eq-Schroedinger} at energy $z$, and therefore
the multiplicity of $z$ as eigenvalue of $H^N_{\hat{Z},Z}$.

\vspace{.2cm}

Given \eqref{eq-Lagdyn}, but also its own sake, it is natural to introduce the
transfer matrices over several sites by
\begin{equation}
\label{eq-iterate} \Tt^z(n,m)\;=\; \Tt_n^z\cdot\ldots\cdot\Tt^z_{m+1}\;,\qquad
n>m \;,
\end{equation}
as well as $\Tt^z(n,n)={\bf 1}$ and $\Tt^z(n,m)=\Tt^z(m,n)^{-1}$ for $n<m$.
With this notation, the solution of the eigenvalue equation
\eqref{eq-Schroedinger} satisfies $\Phi^z_n=\Tt^z(n,m)\Phi_m^z$ and, in
particular, $\Phi^z_n=\Tt^z(n,0)\hat{\Phi}_{\hat{Z}}$. In the following
proposition, it is shown that not only the homogeneous eigenvalue equation, but
also its inhomogeneous counterpart always has a solution for non-real spectral
parameter $z$.

\begin{proposi}
\label{prop-solutions} Given $\psi=(\psi_n)_{n=1,\ldots,N}$ with
$\psi_n\in\mbox{\rm Mat}(L\times L,\CC)$, $\hat{Z},Z\in\overline{\UM_L}$ and
$\Im m(z)>0$, the equation
\begin{equation} \label{eq-inhomogeneous}
(H^N_{\hat{Z},Z}-z\,\one)\,\phi\;=\;\psi\;,
\end{equation}
has a unique solution $\phi=(\phi_n)_{n=1,\ldots,N}$ with $\phi_n\in\mbox{\rm
Mat}(L\times L,\CC)$. It is given by
\begin{equation} \label{eq-solution}
\left(
\begin{array}{c}
T_{n+1}\, \phi^{z}_{n+1}\\
\phi^{z}_n
\end{array}
\right)\;=\; \left(
\begin{array}{cc}
{\bf 1} & \delta_{n=N}\,Z \\
\nul & \one
\end{array}
\right) \left[ \;\sum_{k=1}^n\,\Tt^z(n,k) \left(
\begin{array}{c}
\psi_k \\ \nul
\end{array}
\right) + \Tt^z(n,0) \left(
\begin{array}{cc}
{\bf 1} & \nul \\
-\,\hat{Z} & \one
\end{array}
\right)\left(
\begin{array}{c}
\phi_1 \\ \nul
\end{array}
\right) \right] \;,
\end{equation}
where $n=1,\ldots,N$, together with the constraint {\rm (}which is satisfied
precisely for the solution $\phi${\rm )}
\begin{equation}
\label{eq-constraint}
\phi_{N+1}\;=\;0\;.
\end{equation}
\end{proposi}

\noindent {\bf Proof.} First of all, \eqref{eq-inhomogeneous} is equivalent to
$$
\delta_{n\neq N}
T_{n+1}\phi_{n+1}\,+\,(V_n-\delta_{n=N}Z-\delta_{n=1}\hat{Z})\phi_n
\,+\,\delta_{n\neq 1}T_{n}^*\phi_{n-1}\;=\;\psi_n\;,  \qquad n=1,\ldots,N\;.
$$
Hence, for $n\leq N-1$,
$$
\left(
\begin{array}{c}
T_{n+1}\, \phi^{z}_{n+1}\\
\phi^{z}_n
\end{array}
\right)\;=\; \Tt^z_n\, \left(
\begin{array}{c}
T_{1}\, \phi^{z}_{n}\\
\delta_{n\neq 1}\phi^{z}_{n-1}
\end{array}
\right) \;+\; \left(
\begin{array}{c}
\psi_{n}\\
0
\end{array}
\right) \;,
$$
which by iteration and analyzing the case $n=N$ leads to \eqref{eq-solution}
and the constraint \eqref{eq-constraint}.

In order to show that the constraint can be satisfied, one needs to show that
there is a unique $\phi_1$ such that
$$
\left(
\begin{array}{c}
\one\\ \nul
\end{array}
\right)^* \left(
\begin{array}{cc}
{\bf 1} & Z \\
\nul & \one
\end{array}
\right) \left[ \,\sum_{k=1}^N\;\Tt^z(N,k) \left(
\begin{array}{c}
\one \\ \nul
\end{array}
\right)\,\psi_k + \Tt^z(N,0) \left(
\begin{array}{cc}
{\bf 1} & \nul \\
-\,\hat{Z} & \one
\end{array}
\right)\left(
\begin{array}{c}
\one \\ \nul
\end{array}
\right)\,\phi_1\; \right] \;=\;\nul\;.
$$
For this purpose, let us verify that
$$
\left(
\begin{array}{c}
\one\\ \nul
\end{array}
\right)^* \left(
\begin{array}{cc}
{\bf 1} & Z \\
\nul & \one
\end{array}
\right) \Tt^z(N,0) \left(
\begin{array}{cc}
{\bf 1} & \nul \\
-\,\hat{Z} & \one
\end{array}
\right)\left(
\begin{array}{c}
\one \\ \nul
\end{array}
\right)
$$
is invertible, then
\begin{equation}
\label{eq-phi1} \phi_1\;=\; -\;\left[ \left(
\begin{array}{c}
\one\\ Z
\end{array}
\right)^*  \Tt^z(N,0) \left(
\begin{array}{c}
{\bf 1}  \\
-\,\hat{Z}
\end{array}
\right) \right]^{-1}\,\sum_{k=1}^N\, \left(
\begin{array}{c}
\one\\ Z
\end{array}
\right)^* \Tt^z(N,k) \left(
\begin{array}{c}
\one \\ \nul
\end{array}
\right)\,\psi_k \;,
\end{equation}
leads to a solution which satisfies the constraint. For the proof of the
invertibility, let us set
$$
\left(
\begin{array}{c}
\alpha_n \\ \beta_n
\end{array}
\right)\;=\;  \Tt^z(n,0) \left(
\begin{array}{cc}
{\bf 1} & \nul \\
-\,\hat{Z} & \one
\end{array}
\right)\left(
\begin{array}{c}
\one \\ \nul
\end{array}
\right) \;,
$$
and show by induction that $\alpha_n$ and $\beta_n$ are invertible and that
$\alpha_n(\beta_n)^{-1}\in\UM_L$. At several reprises it will be used that any
$Z\in\UM_L$ is invertible and that $-Z^{-1}\in\UM_L$. One has
$\alpha_1=z\,\one+\hat{Z}-V_1$ and $\beta_1=\one$. Thus all three statements of
the induction hold for $n=1$. Next the definition shows that
$\alpha_n=(z\,\one-V_n)\alpha_{n-1}-\beta_{n-1}$ and $\beta_n=\alpha_{n-1}$ so
that $\alpha_n(\beta_n)^{-1}=z\,\one-V_n-
(\alpha_{n-1}(\beta_{n-1})^{-1})^{-1}$. This implies that $\beta_n$ is
invertible, $\alpha_n(\beta_n)^{-1}\in\UM_L$ and finally that $\alpha_n$ is
therefore also invertible. In the last step, one includes $Z$ in $V_N-Z$ in
order to complete the proof of the invertibility. \hfill $\Box$

\vspace{.2cm}

The proposition clearly implies that $H^N_{\hat{Z},Z}-z\,\one$ is invertible.
As an application, let us calculate the matrix elements of the resolvent. Let
$\pi_n:\CC^L\to \CC^{NL}$ for $n=1,\ldots,N$ denote the partial isometry
$$
\pi_n|l\rangle\;=\;|n,l\rangle\;,\qquad l=1,\ldots,L\;,
$$
where the Dirac notation for localized states in $\CC^{N}\otimes\CC^{L}$ is
used. Then the $L\times L$ Green's matrix is given by
$$
G^z_N(\hat{Z},Z,n,m) \;=\;\pi_n^* (H^N_{\hat{Z},Z}-z\,\one)^{-1}\pi_m\;.
$$
If $\hat{Z}=0$ or $Z=0$ the corresponding argument will be dropped. Of crucial
importance below will be $G^z_N(Z,1,1)$, which will be denoted by $G^z_N(Z)$.
The Green matrices will be calculated in terms of the entries of the transfer
matrix:
\begin{equation}
\label{eq-entrydef} \Tt^z(N,0) \;=\; \left(
\begin{array}{cc} A^z_N & B^z_N \\ C^z_N & D^z_N
\end{array}
\right) \;,
\end{equation}
where all entries are $L\times L$ matrices. These matrices will intervene in
many of the results below. Let us point out that $\Tt^z(N,0)$ and all its
entries do not depend on the boundary conditions $\hat{Z}$ and $Z$.

\vspace{.2cm}

\begin{proposi}
\label{prop-Greenformulas} For Dirichlet boundary condition $\hat{Z}=Z=0$, one
has

\vspace{.1cm}

\noindent {\rm (i)} $\;\;\;\;G^{z}_{N}(1,1)\;=\;(A^z_N)^{-1}B^z_N\;$

\vspace{.1cm}

\noindent {\rm (ii)} $\;\;\;G^{z}_{N}(N,N)\;=\;-\,(A^z_N)^{-1}\;$

\vspace{.1cm}

\noindent {\rm (iii)} $\;\;G^{z}_{N}(1,N)\;=\;-\,C^z_N(A^z_N)^{-1}\;$

\vspace{.1cm}

\noindent {\rm (iv)}
$\;\;G^{z}_{N}(N,1)\;=\;-\,D^z_N\,+\,C^z_N(A^z_N)^{-1}B^z_N$

\end{proposi}

\noindent {\bf Proof.} (i) and (iv) are given in terms of the solution $\phi$
of \eqref{eq-inhomogeneous} with $\psi_n=\delta_{n=1}\one$ by
$G^{z}_{N}(1,1)=\phi_1$ and $G^{z}_{N}(N,1)=\phi_N$. By \eqref{eq-phi1},
$$
G^{z}_{N}(1,1)\;=\;-\;(A^z_N)^{-1} \left(
\begin{array}{c}
\one\\ \nul
\end{array}
\right)^* \Tt^z(N,1) \left(
\begin{array}{c}
\one \\ \nul
\end{array}
\right) \;=\;-\;(A^z_N)^{-1} \left(
\begin{array}{c}
\one\\ \nul
\end{array}
\right)^* \Tt^z(N,0) \left(
\begin{array}{c}
\nul \\ -\,\one
\end{array}
\right) \;,
$$
which is (i). Moreover, (iv) then follows from \eqref{eq-solution}:
$$
G^{z}_{N}(N,1)\;=\; \left(
\begin{array}{c}
\nul \\ \one
\end{array}
\right)^* \left[ \Tt^z(N,1) \left(
\begin{array}{c}
\one \\ \nul
\end{array}
\right) \,+\, \Tt^z(N,0) \left(
\begin{array}{c}
\one \\ \nul
\end{array}
\right)(A^z_N)^{-1}B^z_N\,\right]\;.
$$
For (ii) and (iii), we choose $\psi_n=\delta_{n=N}\one$. Then the solution
$\phi$ gives $G^z_N(1,N)=\phi_1$ and $G^z_N(N,N)=\phi_N$. These are again
readily deduced from \eqref{eq-phi1} and then \eqref{eq-solution}. \hfill
$\Box$

\vspace{.2cm}

Also other combinations of the matrix entries of $\Tt^z(N,0)$ are Green's
functions, {\it e.g.}
$$
G^{z}_{N-1}(1,1)\;=\;(C^z_N)^{-1}D^z_N\;,\qquad
G^{z}_{N-1}(1,N-1)\;=\;-\,(T_NC^z_N)^{-1}\;.
$$
From this and Proposition~\ref{prop-Greenformulas} one can derive norm
estimates on the entries of the transfer matrix, {\it e.g.}
$\|(C^z_N)^{-1}\|\leq (\Im m(z))^{-1}\|T_N\|$ and $\|C^z_N\|\leq (\Im
m(z))^{-1}\|A^z_N\|$. The first of these inequalities can, of course, be
considerably improved by a Combes-Thomas-type estimate. As shown by
\eqref{eq-matrix}, other boundary conditions $\hat{Z},Z$ can be incorporated in
$V_1,V_N$ and then the resulting Hamiltonian has Dirichlet boundary conditions.
The corresponding transfer matrix from $1$ to $N$ is then:
\begin{equation}
\label{eq-boundaryincorp}
\left(
\begin{array}{cc}
\one & Z \\
\nul & \one
\end{array}
\right)
\Tt^z(N,0)
\left(
\begin{array}{cc}
\one & \nul \\
-\,\hat{Z} & \one
\end{array}
\right)
\;=\;
\left(
\begin{array}{cc}
A^z_N - B^z_N\hat{Z} + Z C^z_N - Z D^z_N\hat{Z} & B^z_N +Z D^z_N \\
C^z_N - D^z_N\hat{Z} & D^z_N
\end{array}
\right)
\;.
\end{equation}
From this identity and Proposition~\ref{prop-Greenformulas} one can read off
expressions for Green's matrices with arbitrary boundary conditions. The main
object of the Weyl theory will be the Green matrix with right boundary
condition $Z$ and Dirichlet conditions on the left boundary. This matrix plays
the role of the Weyl-Titchmarch matrix for Sturm-Liouville operators.

\begin{proposi}
\label{prop-Greenformulas2} Set $G^z_N(Z)=G^z_N(0,Z,1,1)$ for
$Z\in\overline{\UM_L}$. Then
\begin{equation}
G^z_N(Z) \; = \; (A^z_N+Z C^z_N)^{-1}(B^z_N +Z D^z_N)  \label{eq-Greendefi} \;
= \; ((D^{\overline{z}}_N)^*Z+(B^{\overline{z}}_N)^*)\,
((C^{\overline{z}}_N)^*Z+(A^{\overline{z}}_N)^*)^{-1} \;.
\end{equation}
\end{proposi}

\vspace{.2cm}

\noindent {\bf Proof.} The first formula follows directly from
Proposition~\ref{prop-Greenformulas}(i) and \eqref{eq-boundaryincorp}. Using
the inverse M\"obius transformation (as defined in the appendix), one can
rewrite it as $G^z_N(Z)= -\,(-Z):\Tt^z(N,0)$. Due to
Proposition~\ref{prop-Moebinv} in the appendix, it follows that $G^z_N(Z)=
-\,\Tt^z(N,0)^{-1}\cdot(-Z)$. Hence we need to calculate the matrix inverse
$\Tt^z(N,0)^{-1}$. As one readily checks that
\begin{equation}
\label{eq-transferinv}
(\Tt^z_n)^{-1}\;=\;\Jj(\Tt^{\overline{z}}_n)^*\Jj^*\;,\qquad \Jj\;=\; \left(
\begin{array}{cc}
\nul & -\,\one \\ \one & \nul \end{array} \right) \;,
\end{equation}
it follows from \eqref{eq-iterate} that
\begin{equation}
\label{eq-transferinv2} \Tt^z(N,0)^{-1}\;=\;
\Jj^*\Tt^{\overline{z}}(N,0)^*\Jj\;=\; \left(
\begin{array}{cc}
(D^{\overline{z}}_N)^* & -\,(B^{\overline{z}}_N)^* \\
-\,(C^{\overline{z}}_N)^* & (A^{\overline{z}}_N)^*
\end{array}
\right) \;.
\end{equation}
From the definition of the M\"obius transformation now follows the second
identity. \hfill $\Box$

\vspace{.2cm}

The entries of the transfer matrices are related to a particular matricial
solution $\psi_n^z=( \psi^{\mbox{\rm\tiny D},z}_n\;\psi^{\mbox{\rm\tiny
A},z}_n)$ of \eqref{eq-Schroedinger}, which are all defined by
\begin{equation}
\label{eq-DAlink} \Tt^z(n,0)\;=\;\Psi^z_n\;=\; \bigl( \Psi^{\mbox{\rm\tiny
D},z}_n\;\Psi^{\mbox{\rm\tiny A},z}_n \bigr) \;=\; \left(\begin{array}{c}
T_{n+1}\psi^{z}_{n+1} \\ \psi^z_n
\end{array}
\right)\;=\;\left(\begin{array}{cc} T_{n+1}\psi^{\mbox{\rm\tiny D},z}_{n+1} &
T_{n+1}\psi^{\mbox{\rm\tiny A},z}_{n+1}
\\
\psi^{\mbox{\rm\tiny D},z}_n & \psi^{\mbox{\rm\tiny A},z}_n
\end{array}
\right) \;.
\end{equation}
The solution $\psi^{\mbox{\rm\tiny D},z}_n$ or $\Psi^{\mbox{\rm\tiny D},z}_n$
is called the Dirichlet solution because it satisfies the Dirichlet condition
on the l.h.s. of \eqref{eq-boundary}. Furthermore $\psi^{\mbox{\rm\tiny
A},z}_n$ or $\Psi^{\mbox{\rm\tiny A},z}_n$ is called the anti-Dirichlet
solution, because the initial condition is orthogonal to the Dirichlet boundary
condition. In the literature \cite{DPS}, $\psi^{\mbox{\rm\tiny D},z}_n$ and
$\psi^{\mbox{\rm\tiny A},z}_n$ are also referred to as matricial orthogonal
polynomials of first and second kind. Any matricial solution of
\eqref{eq-Schroedinger} can be written as combination of the Dirichlet and
anti-Dirichlet solution.

\vspace{.2cm}

\section{Wronskian identities}

For a $2L\times p$ matrix $\Phi$ and a $2L\times p'$ matrix $\Psi$, $1\leq
p,p'\leq 2L$, their Wronskian is defined as
\begin{equation}
\label{eq-Wdef} \Ww(\Phi,\Psi)\;=\;\frac{1}{\imath}\,\Phi^*\Jj\,\Psi\;, \qquad
{\cal J} \;=\; \left(
\begin{array}{cc}
\nul & -\,\one \\
\one & \nul
\end{array}
\right) \;.
\end{equation}
In order to incorporate also the boundary condition $Z$, we also consider the
Wronskians w.r.t.
$$
\Jj(Z) \;=\; \left(
\begin{array}{cc}
\nul & -\,\one \\
\one & Z-Z^*
\end{array}
\right) \;.
$$
In this section, the main focus will be on the Wronskian of the transfer
matrix.

\begin{proposi}
\label{prop-basicWronskian} For $z,\zeta\in\CC$, one has
\begin{eqnarray}
\Tt^z(N,0)^*\Jj(Z)\Tt^\zeta(N,0) & = & \Jj\;+\; (\zeta-\overline{z})\,
\sum_{n=0}^{N-1} \Tt^z(n,0)^*\left(
\begin{array}{cc}
(T_{n+1}T_{n+1}^*)^{-1} & \nul \\
\nul & \nul
\end{array}
\right)\Tt^\zeta(n,0) \nonumber \\
& & \label{eq-basicWron}\\
& & + \;\; \Tt^z(N-1,0)^*\left(
\begin{array}{cc}
(T_{N}^*)^{-1}(Z-Z^*)(T_{N})^{-1} & \nul \\
\nul & \nul
\end{array}
\right)\Tt^\zeta(N-1,0) \;.\nonumber
\end{eqnarray}
Furthermore, let $\Phi^z_n=\Tt^z(n,0)\Phi^z_0$ and
$\Psi^\zeta_n=\Tt^\zeta(n,0)\Psi^\zeta_0$ be $2L\times p$ and $2L\times p'$
matrices and associate $\phi^z_n$ and $\psi^\zeta_n$ as in {\rm
\eqref{eq-solcorr}}. Then
$$
(\Phi^z_N)^*\Jj(Z)\Psi^\zeta_N\;=\; (\Phi^z_0)^*\Jj\Psi^\zeta_0 \;+\;
(\zeta-\overline{z}) \,\langle\phi^z|\psi^\zeta\rangle_N\;+\;
(\phi^z_N)^*(Z-Z^*)\psi^\zeta_N\;,
$$
where
$$
\langle\phi^z|\psi^\zeta\rangle_N\;=\; \sum_{n=1}^N (\phi^z_n)^*\psi^\zeta_n
\;.
$$
\end{proposi}

\noindent {\bf Proof.} One verifies
$$
(\Tt_n^z)^*\Jj(Z)\,\Tt_n^\zeta \;=\; \Jj\;+\; (\zeta-\overline{z}) \, \left(
\begin{array}{cc}
(T_{n}T_n^*)^{-1} & \nul \\
\nul & \nul
\end{array}
\right) \;+\;\left(
\begin{array}{cc}
(T_{n}^*)^{-1}(Z-Z^*)(T_n^*)^{-1} & \nul \\
\nul & \nul
\end{array}
\right)\;.
$$
Iteration then gives \eqref{eq-basicWron}. Evaluation on $\Phi^z_0$ and
$\Psi^\zeta_0$ leads to the second claim. \hfill $\Box$

\vspace{.2cm}

Of central importance will be the case $\zeta=z$.

\begin{proposi}
\label{prop-Qproperties} Let us set $\Qq^z_N=\Ww(\Tt^z(N,0),\Tt^z(N,0))$ and
more generally
\begin{equation}
\label{eq-Qdef} \Qq^z_N(Z)\;=\; \frac{1}{\imath}\,\Tt^z(N,0)^*\Jj(Z)\Tt^z(N,0)
\;.
\end{equation}

\vspace{.1cm}

\noindent {\rm (i)} $\Qq^z_N(Z)^*=\Qq^z_N(Z)$ and $\Qq^z_N(Z+\xi)=\Qq^z_N(Z)$
for all $\xi\in\mbox{\rm Her}(L,\CC)$.

\vspace{.1cm}

\noindent {\rm (ii)} In terms of Dirichlet and anti-Dirichlet solutions, one
has
$$
\Qq^z_N\;=\;\frac{1}{\imath}\,\Jj\;+\; 2\,\Im m(z)\,\langle
\psi^z|\psi^z\rangle_N\;=\;\frac{1}{\imath}\,\Jj\;+\; 2\,\Im m(z)\, \left(
\begin{array}{cc}
\langle\psi^{\mbox{\rm\tiny D},z}|\psi^{\mbox{\rm\tiny D},z}\rangle_N &
\langle\psi^{\mbox{\rm\tiny D},z}|\psi^{\mbox{\rm\tiny A},z}\rangle_N \\
\langle\psi^{\mbox{\rm\tiny A},z}|\psi^{\mbox{\rm\tiny D},z}\rangle_N &
\langle\psi^{\mbox{\rm\tiny A},z}|\psi^{\mbox{\rm\tiny A},z}\rangle_N
\end{array} \right)
$$
\indent and
$$
\Qq^z_N(Z)\;=\;\Qq^z_N\;+\;(\psi^z_N)^* \; \frac{1}{\imath}\;(Z-Z^*)\; \psi^z_N
\;.
$$

In particular, $\Qq^z_N(Z)-\Qq^z_N\geq \nul$ for $Z\in\overline{\UM_L}$.

\vspace{.1cm}

\noindent {\rm (iii)} $\Qq^z_N(Z)^{-1}=\Jj \Qq^{\overline{z}}_N(Z^*)\Jj^*$

\vspace{.1cm}

\noindent {\rm (iv)} Let $\Im m(z)>0$. Then
$$
\Qq^z_{N}\geq \Qq^z_{N-1}\;, \qquad \Qq^z_{N}>\Qq^z_{N-2}\;, \qquad
\Qq^{\overline{z}}_{N}\leq \Qq^{\overline{z}}_{N-1}\;,\qquad
\Qq^{\overline{z}}_{N}<\Qq^{\overline{z}}_{N-2}\;.
$$

\vspace{.1cm}

\noindent {\rm (v)} {\rm signature}$(\Qq^z_N(Z))=(L,L)=\;${\rm
signature}$(\Qq^{\overline{z}}_N(Z^*))$

\vspace{.1cm}

\noindent {\rm (vi)} For $\Im m(z)>0$ and $Z\in\overline{\UM_L}$, the spectrum
of $\Qq^z_N(Z)$ is contained in $(-1,\infty)$ and that of

$\Qq^{\overline{z}}_N(Z^*)$ in $(-\infty,1)$.

\end{proposi}

\vspace{.2cm}

\noindent {\bf Proof.} Item (i) follows directly from $\Jj(Z)^*=-\Jj(Z)$. Item
(ii) follows directly from Proposition~\ref{prop-basicWronskian} as well as
$\psi^{\mbox{\rm\tiny D},z}_N=C^z_N$ and $\psi^{\mbox{\rm\tiny A},z}_N=D^z_N$.
For the proof of (iii), we appeal to \eqref{eq-transferinv2}, giving
$$
\Qq^z_N(Z)^{-1} \; = \; \imath\,\Tt^z(N,0)^{-1}\Jj(Z)^{-1}(\Tt^z(N,0)^{-1})^*
\; = \;
\imath\,\Jj^*\Tt^{\overline{z}}(N,0)^*\Jj\Jj(Z)^{-1}\Jj^*\Tt^{\overline{z}}(N,0)\Jj
\;,
$$
which together with $\Jj\Jj(Z)^{-1}\Jj=\Jj(Z^*)$ implies the claimed formula.
Next,
$$
\Qq^z_N\,-\,\Qq^z_{N-1}\;=\; \Tt^z(N,0)^*\left(
\begin{array}{cc}
(T_{N+1}T_{N+1}^*)^{-1} & \nul \\
\nul & \nul
\end{array}
\right)\Tt^z(N,0) \;,
$$
from this follows the first claim of (iv). Furthermore, $\Qq^z_N-\Qq^z_{N-2}$
is equal to
$$
\Tt^z(N,0)^*\left(
\begin{array}{cc}
(T_{N+1}T_{N+1}^*)^{-1} & \nul \\
\nul & \nul
\end{array}
\right)\Tt^z(N,0) + \Tt^z(N-1,0)^*\left(
\begin{array}{cc}
(T_{N}T_{N}^*)^{-1} & \nul \\
\nul & \nul
\end{array}
\right)\Tt^z(N-1,0) \;.
$$
Hence it is sufficient to show that
$$
(\Tt^z_N)^*\left(
\begin{array}{cc}
(T_{N+1}T_{N+1}^*)^{-1} & \nul \\
\nul & \nul
\end{array}
\right)\Tt^z_N \;+ \;\left(
\begin{array}{cc}
(T_{N}T_{N}^*)^{-1} & \nul \\
\nul & \nul
\end{array}
\right) \;>\;0\;.
$$
Because the l.h.s. is clearly positive semi-definite, it is sufficient to
verify that the kernel is trivial. Let hence $\left(
\begin{array}{c} v \\ w \end{array} \right)$ be in the kernel.
Using \eqref{eq-transferinv}, this means
$$
\Jj^*\left(
\begin{array}{cc}
(T_{N+1}T_{N+1}^*)^{-1} & \nul \\
\nul & \nul
\end{array}
\right)\Tt^z_N\left(
\begin{array}{c} v \\ w \end{array} \right) \;=\; \Tt^{\overline{z}}_N\Jj \left(
\begin{array}{cc}
(T_{N}T_{N}^*)^{-1} & \nul \\
\nul & \nul
\end{array}
\right) \left(
\begin{array}{c} v \\ w \end{array} \right)\;,
$$
or
$$
\left(
\begin{array}{cc}
\nul & \nul \\
(T_{N+1}T_{N+1}^*)^{-1}(z\,\one-V_N)(T_N)^{-1} & (T_{N+1}T_{N+1}^*)^{-1} T_N^*
\end{array}
\right) \left(
\begin{array}{c} v \\ w \end{array} \right) \;=\; \left(
\begin{array}{cc}
(T_{N})^{-1} & \nul \\
\nul & \nul
\end{array}
\right) \left(
\begin{array}{c} v \\ w \end{array} \right)\;.
$$
This implies $v=w=0$ because $T_N$ and $T_{N+1}$ are invertible. Finally, item
(v) follows from \eqref{eq-Qdef}, Sylvester's theorem and the fact that the
spectrum of $\frac{1}{\imath}\Jj$ is $\{-1,1\}$ with multiplicity $L$ for each
eigenvalue, so that the signature of $\frac{1}{\imath}\Jj$ is $(L,L)$. Item
(vi) follows from the same fact and (ii). \hfill $\Box$

\section{Isotropic subspaces}
\label{sec-isotropic}

In this section we analyze $L$-dimensional subspaces of $\CC^{2L}$ naturally
associated to the Jacobi matrices $H^N$. The Grassmannian $\GM_L$ of
$L$-dimensional subspaces is by definition the set of equivalence classes of
complex $2L\times L$-matrices  $\Phi$ of rank $L$ w.r.t. to the equivalence
relation $\Phi\sim\Psi$ $\Leftrightarrow$ $\Phi=\Psi c$ for some
$c\in\,$GL$(L,\CC)$. The elements of $\GM_L$ are denoted by $[\Phi]_\sim$. A
subset $\GM_L^\inv\subset \GM_L$ of full measure are those subspaces
represented by a matrix $\Phi=\left(\begin{array}{c} a \\ b
\end{array} \right)$ with invertible $b\in\,$GL$(L,\CC)$. This set is the
domain of the stereographic projection
$$
\pi:\GM_L^\inv\to \mbox{\rm Mat}(L,\CC)\;, \qquad \pi([\Phi]_\sim)\;=\;ab^{-1}
\;, \qquad \Phi\;=\; \left(\begin{array}{c} a
\\ b \end{array} \right)\;.
$$
An $L$-dimensional plane $[\Phi]_\sim$ is called hermitian Lagrangian (or
maximally isotropic subspace w.r.t. $\!\Jj$) if and only if $\Phi^*\Jj\Phi=0$
(or equivalently $\Ww(\Phi,\Phi)=\imath(a^*b-b^*a)=0$). The Lagrangian
Grassmannian is denoted by $\LM_L$ and we set $\LM_L^\inv=\LM_L\cap
\GM_L^\inv$. It is proven in \cite{SB} that $\LM_L$ is diffeomorphic to the
unitary group U$(L)$. In connection with the Jacobi matrices we analyze the
plane $\Phi_G=\left(\begin{array}{c} -\,G
\\ \one \end{array} \right)$ for any $G\in \mbox{\rm Mat}(L,\CC)$, defined as
in \eqref{eq-boundaryplanes}.

\vspace{.2cm}

\begin{proposi}
\label{prop-GreenLag} Let $Z\in \overline{\UM_L}$ and $\Im m(z)>0$.

\vspace{.1cm}

\noindent {\rm (i)} For $G=G^z_N(Z)$, one has
\begin{equation}
\label{eq-GreenLag}
\left(
\begin{array}{cc}
\one & Z \\
\nul & \one
\end{array}
\right)\,\Tt^z(N,0)\, \Phi_G\;\in\;\LM_L\;.
\end{equation}

\vspace{.1cm}

\noindent {\rm (ii)} If
\begin{equation}
\label{eq-GreenLagHyp} \left(
\begin{array}{cc}
\one & Z \\
\nul & \one
\end{array}
\right)\,\Tt^z(N,0)\, \Phi_G\;\in\;\LM_L\;,\qquad \Tt^z(N,0)\,
\Phi_G\;\in\;\GM_L^\inv\;,
\end{equation}

then $G=G^z_N(Z+\xi)$ for some $\xi\in\mbox{\rm Her}(L,\CC)$.

\end{proposi}

\vspace{.2cm}

\noindent {\bf Proof.} (i) First let us point out that the plane in
\eqref{eq-GreenLag} is indeed $L$-dimensional. If $G=G^z_N(Z)$, then $G=\phi_1$
where $\phi$ is the solution of \eqref{eq-inhomogeneous} with $\hat{Z}=\nul$
and r.h.s. $\psi_n=\delta_{n=1}\,\one$. The constraint equation
\eqref{eq-constraint} can be expressed using \eqref{eq-solution} (in which the
two terms on the r.h.s. can be resembled) as
$$
\left(
\begin{array}{c}
\one \\
\nul
\end{array}
\right)^* \left(
\begin{array}{cc}
\one & Z \\
\nul & \one
\end{array}
\right)\,\Tt^z(N,0)\, \Phi_G \;=\;\nul\;.
$$
But this means that the plane \eqref{eq-GreenLag} is of the form $\left(
\begin{array}{c} \nul \\ b \end{array} \right) $ with some $b\in\,$GL$(L,\CC)$.
This plane is Lagrangian.

\vspace{.1cm}

The second hypothesis in \eqref{eq-GreenLagHyp} is equivalent to requiring
$C^z_NG-D^z_N$ to be invertible. The first hypothesis means that
$$
\bigr(A^z_NG-Bz_N+Z(C^z_NG-D^z_N)\bigr)^*(C^z_NG-D^z_N) \;=\; (C^z_NG-D^z_N)^*
\bigl(A^z_NG-B^z_N+Z(C^z_NG-D^z_N)\bigr)\;.
$$
Multiplying this from the right by $(C^z_NG-D^z_N)^{-1}$ and from the left by
$((C^z_NG-D^z_N)^*)^{-1}$ shows that
$$
-\,\xi\;=\;(A^z_NG-B^z_N)(C^z_NG-D^z_N)^{-1}\,+\,Z\;
$$
is self-adjoint. Using the M\"obius transformation this equation can be
rewritten as $-\xi=\Tt^z(N,0)\cdot(-G)+Z$. But by
Proposition~\ref{prop-Moebinv} this is equivalent to $G=-
\Tt^z(N,0)^{-1}\cdot(-Z-\xi)$, which by Proposition~\ref{prop-Greenformulas2}
shows $G=G^z_N(Z+\xi)$. \hfill $\Box$

\vspace{.2cm}

Now we shift perspective and interpret Proposition~\ref{prop-GreenLag} in terms
of $\Qq^z_N(Z)$ considered as a quadratic form on $\CC^{2L}$. A plane
$[\Phi]_\sim\in\GM_L$ is called isotropic for $\Qq^z_N(Z)$ if and only if
$\Phi^* \Qq^z_N(Z)\Phi=0$.

\begin{proposi}
\label{prop-isotropic} Let $N\geq 2$.

\vspace{.1cm}

\noindent {\rm (i)} Let $G=G^z_N(Z)$. Then $\Phi_G$ is an isotropic plane for
$\Qq^z_N(Z)$.

\vspace{.1cm}

\noindent {\rm (ii)} Suppose that $[\Phi]_\sim\in\GM_L$ is an isotropic plane
for $\Qq^z_N(Z)$. Then $[\Phi]_\sim$ is represented by $\Phi_G$

with $G\in\UM_L$ and is, in particular, in $\GM^\inv_L$. If, moreover,
$\Tt^z(N,0)\Phi_G\in\GM_L^\inv$, then

$G=G^z_N(Z+\xi)$ for some $\xi\in\mbox{\rm Her}(L,\CC)$.

\end{proposi}

\vspace{.2cm}

\noindent {\bf Proof.} (i) is just a reformulation of
Proposition~\ref{prop-GreenLag}(i). For (ii) one first notes that by
Proposition~\ref{prop-Qproperties} or directly \eqref{eq-basicWron} one has
$\Qq^z_N(Z)=-\imath\, \Jj+\Pp$ with $\Pp>\nul$. Using the hypothesis, it hence
follows that $\Ww(\Phi,\Phi)=-\Phi^*\Pp\Phi<\nul$. Therefore
$[\Phi]_\sim\in\GM_L^\inv$ and $\pi([\Phi]_\sim)\in (-\UM_L)$ by
Lemma~\ref{lem-Wronskinv} just below. But this is precisely the first claim
because every $[\Phi]_\sim\in\GM_L^\inv$ is represented by some $\Phi_G$ and
$\pi([\Phi_G]_\sim)=-G$. The second claim then follows from
Proposition~\ref{prop-GreenLag}(ii).\hfill $\Box$

\vspace{.2cm}

\begin{lemma}
\label{lem-Wronskinv} Suppose that $[\Phi]_\sim\in\GM_L$ satisfies
$\Ww(\Phi,\Phi)>\nul$. Then $[\Phi]_\sim\in\GM_L^\inv$ and $\pi([\Phi]_\sim)\in
\UM_L$. If $\Ww(\Phi,\Phi)<\nul$, then $-\pi([\Phi]_\sim)\in \UM_L$.
\end{lemma}

\vspace{.2cm}

\noindent {\bf Proof.} Let $\Phi=\left(
\begin{array}{c} a \\ b \end{array} \right)$. Suppose $b$ is not invertible,
that is, there exists a non-vanishing $v\in\CC^L$ such that $bv=0$. Then
$v^*\Ww(\Phi,\Phi)v=\imath v^*(a^*b-b^*a)v=0$, in contradiction to the supposed
positivity. Similarly, $a$ is invertible. Then
$$
\Ww(\Phi,\Phi)\;=\;\frac{1}{\imath}\;(-a^*b+b^*a)\;=\;\frac{1}{\imath}\;b^*
\bigl(ab^{-1} -(ab^{-1})^*\bigr)b\;.
$$
This implies the second claim. \hfill $\Box$

\vspace{.2cm}

\section{The Weyl surface and the Weyl disc}

By Proposition~\ref{prop-isotropic} the isotropic planes of $\Qq_N^z(Z)$ are in
the domain of the stereographic projection $\pi$. Hence the following
definition is possible.

\begin{defini}
\label{def-Weyldisc} Let $\Im m(z)\neq 0$. Associated to a Jacobi matrix $H^N$
with matrix entries are the Weyl surface
\begin{equation}
\label{eq-Weylsurface} \partial_L\WM_N^z\;=\; -\,\pi\left(\,\left\{\,
[\Phi]_\sim\in\GM_L
 \;\left|\;\Phi\;\,\mbox{\rm isotropic for }\Qq_N^z
 \;\right.\right\}\;\right)\,,
\end{equation}
and the closed Weyl disc
\begin{equation}
\label{eq-Weyldisc} \overline{\WM_N^z}\;=\; -\,\pi\left(\,\left\{\,
[\Phi]_\sim\in\GM_L
 \;\left|\;\Phi\;\,\mbox{\rm isotropic for }\Qq_N^z(\sigma^z Z) \;\;\mbox{\rm for some
 }Z\in\overline{\UM_L} \;\right.\right\}\;\right)\,,
\end{equation}
where $\sigma^z$ is the sign of $\Im m(z)$. The interior of
$\;\overline{\WM^z_N}$ is Weyl disc ${\WM_N^z}$.
\end{defini}

\begin{proposi}
\label{prop-Weylineq} One has
\begin{equation}
\label{eq-Weylrewrite}
\partial_L \WM_N^z\;=\;
\left\{\,G\in\,\mbox{\rm Mat}(L,\CC)\; \left|\; \Phi_G^*\Qq_N^z\Phi_G=\nul
\;\right.\right\} \;.
\end{equation}
For $\Im m(z)>0$,
\begin{equation}
\label{eq-Weylrewrite2} \overline{\WM_N^z}\;=\; \left\{\,G\in\,\mbox{\rm
Mat}(L,\CC)\; \left|\; \Phi_G^*\Qq_N^z\Phi_G\leq\nul \;\right.\right\} \;,
\end{equation}
and hence
\begin{equation}
\label{eq-Weylrewrite3} {\WM_N^z}\;=\; \left\{\,G\in\,\mbox{\rm Mat}(L,\CC)\;
\left|\; \Phi_G^*\Qq_N^z\Phi_G<\nul \;\right.\right\} \;.
\end{equation}
These are all subsets of the upper half-plane $\UM_L$. For $\Im m(z)<0$, the
inequalities in {\rm \eqref{eq-Weylrewrite2}} and {\rm \eqref{eq-Weylrewrite3}}
are reversed and the closed Weyl disc lies in the lower half-plane.
\end{proposi}

\noindent {\bf Proof.} By Proposition~\ref{prop-isotropic}(ii) every isotropic
plane of $\Qq_N^z$ is represented by some $\Phi_G$. As
$\pi([\Phi_G]_\sim)=-\,G$ the equality \eqref{eq-Weylrewrite} follows. For the
proof of \eqref{eq-Weylrewrite2}, let us first note that
$\Qq_N^z(Z)\geq\Qq_N^z$ by Proposition~\ref{prop-Qproperties}(ii). Hence, if
$\Phi$ is isotropic for $\Qq_N^z(Z)$, then $\Phi^*\Qq^z_N\Phi\leq \nul$. Again
as all isotropic planes of $\Qq^z_N(Z)$ are represented by some $\Phi_G$, the
inclusion $\subset$ of \eqref{eq-Weylrewrite2} follows. For the converse, given
$\Phi_G^*\Qq^z_N\Phi_G\leq \nul$, we have to find $Z\in\overline{\UM_L}$ such
that $\Phi_G^*\Qq^z_N(Z)\Phi_G= \nul$, that is
$$
\Phi_G^*\Qq^z_N\Phi_G\,+\,(C^z_NG-D^z_N)^*\,
\imath(Z^*-Z)\,(C^z_NG-D^z_N)\;=\;\nul \;.
$$
If $C^z_NG-D^z_N$ is invertible, this equation is readily solved (of course,
only the imaginary part of $Z$ is uniquely determined). If $v\in\CC^L$ is in
the kernel of $C^z_NG-D^z_N$, then
\begin{equation} \label{eq-Wrcalc}
\Phi_G^*\Qq^z_N\Phi_G\;=\;
(A^z_NG-B^z_N)^*(C^z_NG-D^z_N)-(C^z_NG-D^z_N)^*(A^z_NG-B^z_N) \;.
\end{equation}
shows that $v^*\Phi_G^*\Qq^z_N\Phi_Gv=0$ so that $v$ is in the kernel of
$\Phi_G^*\Qq^z_N\Phi_G$. Therefore, the following lemma completes the proof of
{\rm \eqref{eq-Weylrewrite2}}. From this also follows {\rm
\eqref{eq-Weylrewrite3}}. The case $\Im m(z)<0$ is dealt with in a similar
manner. \hfill $\Box$

\vspace{.2cm}

\begin{lemma}
\label{lem-solve} Let $A,B\in\mbox{\rm Mat}(L,\CC)$ satisfy $A\geq \nul$ and
suppose that $\mbox{\rm ker}(B)\subset\mbox{\rm ker}(A)$. Then the equation
$A=B^*YB$ has a solution $Y\geq\nul$.
\end{lemma}

\vspace{.2cm}

\noindent {\bf Proof.} Let $P$ be the projection on the positive spectrum of
$A$, {\it i.e.} $A=PAP$. Then $Av\neq 0$ is equivalent to $Pv\neq 0$, which
implies $P(Pv)\neq 0$, so by hypothesis $BPv\neq 0$. Let $Q$ be the projection
on the span of $BV$. Then dim$(Q)=\,$dim$(P)$ and $QBP:P\CC^L\to Q\CC^L$ is
invertible. Set $Y=QYQ=((QBP)^{-1})^*PAP(QBP)^{-1}$. \hfill $\Box$

\vspace{.2cm}

The notation $\partial_L\WM_N^z$ reflects that the Weyl surface is the maximal
boundary of $\partial\WM_N^z=\overline{\WM_N^z}\,/\,\WM^z_N$. It is possible to
define other strata of $\partial\WM_N^z$, but they will not be used here. In
order to realize that the Weyl disc and surface merit their names let us next
define the following objects.

\begin{defini}
\label{def-Weyobjects} Let $\Im m(z)\neq 0$. Associated to a Jacobi matrix
$H^N$ with matrix entries are the radial and center operators
\begin{equation}
\label{eq-radius} R^z_N\;=\; \left[\,\left(
\begin{array}{c}
\one \\
\nul
\end{array}
\right)^* \Qq^z_N \left(
\begin{array}{c}
\one \\
\nul
\end{array}
\right)\right]^{-1}\;,\qquad S^z_N\;=\; R^z_N\,\left(
\begin{array}{c}
\one \\
\nul
\end{array}
\right)^* \Qq^z_N \left(
\begin{array}{c}
\nul \\
\one
\end{array}
\right) \;.
\end{equation}
\end{defini}

\vspace{.2cm}

Here the inverse in the definition of $R^z_N$ exists because of
Proposition~\ref{prop-Qproperties}. Using the definitions \eqref{eq-Qdef} and
\eqref{eq-entrydef}, one obtains
\begin{equation}
\label{eq-radius2} R^z_N\;=\; \imath\;\left[ (C^z_N)^*A^z_N-(A^z_N)^*C^z_N
\right]^{-1}  \;, \qquad S^z_N \;=\;{\imath}\;R^z_N\; \left[
(A^z_N)^*D^z_N-(C^z_N)^*B^z_N \right] \;.
\end{equation}
The following proposition recollects a few basic properties of these objects.

\begin{proposi}
\label{prop-basicformulas} Let $N\geq 2$.

\vspace{.1cm}

\noindent {\rm (i)} The following Wronskian identities hold:
$$
R^z_N\;=\; \left[\, 2\;\Im m(z)\;\langle \psi^{\mbox{\rm\tiny D},z}|
\psi^{\mbox{\rm\tiny D},z}\rangle_N \,\right]^{-1} \;, \qquad S^z_N \;=\;
\imath\;R^z_N \left[\,\one-(z-\overline{z})\,\langle \psi^{\mbox{\rm\tiny
D},z}| \psi^{\mbox{\rm\tiny A},z}\rangle_N \,\right] \;.
$$

In particular, for $\Im m(z)>0$ one has $R^z_N>\nul$ and
$(-R^{\overline{z}}_N)>\nul$.

\vspace{.1cm}

\noindent {\rm (ii)} Let $\Im m(z)>0$. Then $R^z_{N-1}\geq R^z_{N}$ and
$R^z_{N-2}> R^z_{N}$ as well as $R^{\overline{z}}_{N-1}\leq
R^{\overline{z}}_{N}$ and $R^{\overline{z}}_{N-2}< R^{\overline{z}}_{N}$.

\vspace{.1cm}

\noindent {\rm (iii)} One has
$$
(S^z_N)^*=S^{\overline{z}}_N\;,
\qquad
R^{\overline{z}}_N\;=\;\left(
\begin{array}{c}
\nul \\
\one
\end{array}
\right)^* \Qq^z_N \left(
\begin{array}{c}
\nul \\
\one
\end{array}
\right) \,-\,(S^z_N)^*(R^z_N)^{-1}S^z_N\;.
$$

\vspace{.1cm}

\noindent {\rm (iv)} If $H^N$ is real, $R^{\overline{z}}_N=-\overline{R^z_N}$
and $S^{\overline{z}}_N=\overline{S^z_N}$.

\end{proposi}

\noindent {\bf Proof.} (i) The identities follow directly from
Proposition~\ref{prop-Qproperties}(ii). The positivity of $R^z_N$ and
$-R^{\overline{z}}_N$ follows from the fact that $\langle \psi^{\mbox{\rm\tiny
D},z}| \psi^{\mbox{\rm\tiny D},z}\rangle_N>\nul$ and $\langle
\psi^{\mbox{\rm\tiny D},\overline{z}}| \psi^{\mbox{\rm\tiny
D},\overline{z}}\rangle_N>\nul$ for $N>1$.

\vspace{.1cm}

(ii) Proposition~\ref{prop-Qproperties}(iv) implies first of all $\langle
\psi^{\mbox{\rm\tiny D},z}| \psi^{\mbox{\rm\tiny D},z}\rangle_N\geq \langle
\psi^{\mbox{\rm\tiny D},z}| \psi^{\mbox{\rm\tiny D},z}\rangle_{N-1}$, and
second of all $\langle \psi^{\mbox{\rm\tiny D},z}| \psi^{\mbox{\rm\tiny
D},z}\rangle_N> \langle \psi^{\mbox{\rm\tiny D},z}| \psi^{\mbox{\rm\tiny
D},z}\rangle_{N-2}$. The claim now follows because the function
$t\in\RR_{>0}\mapsto -t^{-1}$ is operator monotonous (if $\nul<A\leq B$, then
$\nul<B^{-1}\leq A^{-1}$).

\vspace{.1cm}

(iii) Let us rewrite the identity of Proposition~\ref{prop-Qproperties}(iii) as
$$
\Jj\;=\; \Qq^{\overline{z}}_N \left(
\begin{array}{c}
\nul \\
\one
\end{array}
\right)\left(
\begin{array}{c}
\one \\
\nul
\end{array}
\right)^* \Qq^z_N \,-\, \Qq^{\overline{z}}_N \left(
\begin{array}{c}
\one \\
\nul
\end{array}
\right)\left(
\begin{array}{c}
\nul \\
\one
\end{array}
\right)^* \Qq^z_N\;.
$$
Writing out the upper left and right entries of this identity in terms of the
radial and center operators gives
$$
\nul\;=\;(R^{\overline{z}}_N)^{-1}S^{\overline{z}}_N(R^z_N)^{-1}\,-\,
(R^{\overline{z}}_N)^{-1}(S^z_N)^*(R^z_N)^{-1}\;,
$$
and
$$
-\,\one\;=\;(R^{\overline{z}}_N)^{-1}S^{\overline{z}}_N(R^z_N)^{-1}S^z_N\,-\,
(R^{\overline{z}}_N)^{-1} \left(
\begin{array}{c}
\nul \\
\one
\end{array}
\right)^* \Qq^z_N \left(
\begin{array}{c}
\nul \\
\one
\end{array}
\right) \;.
$$
These two equations directly lead to the claims.

\vspace{.1cm}

Item (v) follows directly from the definition \eqref{eq-radius2} because
$A^{\overline{z}}_N=\overline{A^{{z}}_N}$ holds for real $H^N$ and similar
identities the other entries of the transfer matrix. \hfill $\Box$

\vspace{.2cm}

\begin{proposi}
\label{prop-Weylsurface} The Weyl surface satisfies
\begin{eqnarray}
\partial_L\WM_N^z
& = & \left\{\,\left.S^z_N+(R^z_N)^{\frac{1}{2}} W
(-R^{\overline{z}}_N)^{\frac{1}{2}} \;\right|\; W\in\,\mbox{\rm U}(L)
\;\right\} \label{eq-Weylsurfacefrom1}
\\
& = & \left\{\,G\in\,\mbox{\rm Mat}(L,\CC)\;\left|\; \left|\;
(R^{{z}}_N)^{-\frac{1}{2}} (G-S^z_N)
(-R^{\overline{z}}_N)^{-\frac{1}{2}}\right|^2\,=\,\one\;\right.\right\}
\nonumber
\\
& = & \left\{\,G\in\,\mbox{\rm Mat}(L,\CC)\; \left|\; 2\,|\Im
m(z)|\;\sum_{n=1}^N\left|\psi^{\mbox{\rm\tiny D},{z}}_nG-\psi^{\mbox{\rm\tiny
A},{z}}_n\right|^2\;=\;\imath\,(G^*-G)\;\right.\right\}\;.
\label{eq-Weylsurfacefrom3}
\end{eqnarray}
The same equalities hold for $\overline{\WM_N^z}$ and $\WM_N^z$ if one allows
$W$ to run through $\overline{\DM_L}$ and $\DM_L$ respectively, and replaces
the two equalities by $\leq$ and $<$ respectively.
\end{proposi}

\noindent {\bf Proof.} Let us write out the equation
$\Phi_G^*\Qq_N^z\Phi_G=\nul$ appearing in \eqref{eq-Weylrewrite} explicitly
using the radial and center operators:
$$
G^* (R^z_N)^{-1}G\,-\, G^*(R^z_N)^{-1}S^z_N \,-\,
(S^z_N)^*(R^z_N)^{-1}G\,-\,\left(
\begin{array}{c}
\one \\
\nul
\end{array}
\right)^* \Qq^z_N \left(
\begin{array}{c}
\one \\
\nul
\end{array}
\right)\;=\;\nul\;.
$$
Completing the square gives
$$
(G-S^z_N)^*(R^z_N)^{-1}(G-S^z_N)\;=\; (S^z_N)^*(R^z_N)^{-1}S^z_N\,-\,\left(
\begin{array}{c}
\one \\
\nul
\end{array}
\right)^* \Qq^z_N \left(
\begin{array}{c}
\one \\
\nul
\end{array}
\right)\;=\;-\,R^{\overline{z}}_N\;,
$$
where the last identity is precisely Proposition~\ref{prop-basicformulas}(iii).
From this follow directly the first two equalities of the proposition. For the
last equality, one has to write out $\Phi_G^*\Qq_N^z\Phi_G=\nul$ using
Proposition~\ref{prop-Qproperties}(ii). The cases of $\overline{\WM_N^z}$ and
$\WM_N^z$ follow by replacing equalities by inequalities. \hfill $\Box$

\vspace{.2cm}

Now we can resume the main results on the finite volume Weyl theory. Item (i)
of the following theorem can be deduced from the results proven by Fukushima
\cite{Fuk}, results related to (iii) are given by Orlov \cite{Orl}.

\begin{theo}
\label{theo-Weyl} Let $\Im m(z)\neq 0$ and $N\geq 2$.

\vspace{.1cm}

\noindent {\rm (i)} Green's matrix $G^z_N(\xi)$ for self-adjoint boundary
conditions $\xi\in\,\mbox{\rm Her}(L,\CC)$ lies on the Weyl

surface $\partial_L\WM^z_N$.

\vspace{.1cm}

\noindent {\rm (ii)} For $\Im m(z)> 0$, the map $Z\mapsto G^z_N(Z)$ establishes
a diffeomorphism between the upper half

plane $\UM_L$ and the Weyl disc $\WM^z_N$. For $\Im m(z)< 0$, the same holds if
$Z$ runs through the lower

half-plane.

\vspace{.1cm}

\noindent {\rm (iii)} The Weyl discs are strictly nested, that is
$$
\WM^z_{N}\subset\WM^z_{N-1}\;, \qquad
\partial\WM^z_{N-1}\cap\overline{\WM^z_{N+1}}\;=\;\emptyset\;.
$$

\end{theo}

\noindent {\bf Proof.} (i) This follows from
Proposition~\ref{prop-isotropic}(i) and \eqref{eq-Weylsurface}. For (ii) one
argues  similarly to verify $G^z_N(Z)\in\WM_N^z$ for $Z\in\UM_L$. To show the
surjectivity, let us note that $\Phi_G^*\Qq_N^z\Phi_G<\nul$ implies by
\eqref{eq-Wrcalc} the invertibility of $C^z_NG-D_N^z$. Again from
\eqref{eq-Wrcalc} one then deduces that
$Z=-(A^z_NG-B_N^z)(C^z_NG-D_N^z)^{-1}=-\Tt^z(N,0)\cdot (-G)$ is in the upper
half-plane. Hence $G=-\Tt^z(N,0)^{-1}\cdot (-Z)$ exists (arguing as in the
proof of Proposition~\ref{prop-Greenformulas2}) and is therefore $G=G^z_N(Z)$
by that proposition. Item (iii) follows from the fact that the l.h.s. in the
equality of \eqref{eq-Weylsurfacefrom3} is increasing in $N$ and strictly
increasing whenever $2$ consecutive terms are taken out, namely
$\sum_{n=N-1}^N\left|\psi^{\mbox{\rm\tiny D},{z}}_nG-\psi^{\mbox{\rm\tiny
A},{z}}_n\right|^2>\nul$. The latter fact follows upon evaluating
Proposition~\ref{prop-Qproperties}(iv) on $\Phi_G$. \hfill $\Box$

\vspace{.2cm}

A first application of the Weyl discs are the following estimates on the finite
volume Green's matrices. They are completely analogous to the scalar case, but
the second one is far from optimal ({\it cf.} \cite{FHS} for a significant
improvement).

\begin{proposi}
\label{prop-Greenestimates}
For $\xi,\xi'\in\,\mbox{\rm Her}(L,\CC)$, one has
$$
\|G^z_N(\xi)-G^z_N(\xi')\|
\;\leq\;\,2\;\sqrt{\|R_N^z\|\|R_N^{\overline{z}}\|}\;.
$$
Moreover, for any $N\geq 2$,
$$
\|R^z_N\|
\;\leq\;
\left[ \,2\;\Im m(z)^2\;\sum_{n=2}^{N}
\,\frac{1}{\|T_n\|}\;\right]^{-1}
\;.
$$
\end{proposi}

\noindent {\bf Proof}. By Theorem~\ref{theo-Weyl} there is a $W\in\,$U$(L)$
such that  for $\xi\in\,$Her$(L,\CC)$ one has $G^z_N(\xi)=S^z_N+
(R^{{z}}_N)^{\frac{1}{2}}W(-R^{\overline{z}}_N)^{\frac{1}{2}}$. Hence
$\|G^z_N(\xi)-S^z_N\|^2\leq \|R^{\overline{z}}_N\|\, \|R^{{z}}_N\|$ as both
$(-R^{\overline{z}}_N)$ and $(R^z_N)^{-1}$ are positive definite. Therefore the
first inequality follows from
$\|G^z_N(\xi)-G^z_N(\xi')\|\leq\|G^z_N(\xi)-S^z_N\|+\|G^z_N(\xi')-S^z_N\|$.

\vspace{.1cm}

For the proof of the second inequality let us first recall the identity
$$
\Ww( \Psi^{\mbox{\rm\tiny D},{z}}_n,\Psi^{\mbox{\rm\tiny D},{z}}_n)=2\,\Im
m(z) \sum_{m=1}^n|\psi^{\mbox{\rm\tiny D},{z}}_m|^2$$ 
following from
Proposition~\ref{prop-basicWronskian} with $\zeta=z$ and $Z=\nul$. Because
$\psi^{\mbox{\rm\tiny D},{z}}_1=\one$ and all the terms on the r.h.s. are
positive semi-definite, it follows that
$$
\imath\, (\psi^{\mbox{\rm\tiny D},{z}}_{n+1})^*T_{n+1}^*\psi^{\mbox{\rm\tiny
D},{z}}_n-\imath\,(\psi^{\mbox{\rm\tiny D},{z}}_n)^*T_{n+1}\psi^{\mbox{\rm\tiny
D},{z}}_{n+1} \;\geq\;\imath\,(\overline{z}-z)\,\one\;.
$$
Therefore, one has for any unit vector $v\in\CC^L$,
$$
\|\psi^{\mbox{\rm\tiny D},{z}}_nv\|\;\|\psi^{\mbox{\rm\tiny D},{z}}_{n+1}v\|
\;\geq\; \frac{\Im m(z)}{\|T_{n+1}\|}\;.
$$
Summing over $n=1,\ldots,N-1$ and using the Cauchy-Schwarz inequality gives
$$
\sum_{n=1}^N\;\|\psi^{\mbox{\rm\tiny D},{z}}_nv\|^2 \;\geq\; \Im m(z)\;
\sum_{n=1}^{N-1}\;\frac{1}{\|T_{n+1}\|}\;.
$$
Now by definition of the radial operator,
$$
\sum_{n=1}^N\;\|\psi^{\mbox{\rm\tiny D},{z}}_nv\|^2 \;=\; \frac{1}{2\,\Im
m(z)}\; \langle v|(R^z_N)^{-1}|v\rangle \;\leq \;\frac{1}{2\,\Im
m(z)}\;\frac{1}{\|R^z_N\|}\;,
$$
which implies the second inequality. \hfill $\Box$

\vspace{.2cm}

Before going on to semi-infinite Jacobi matrices, let us stress once again what
precisely is the r\^ole of the Weyl surface $\partial_L\WM^z_N$. It encodes all
the possible right boundary conditions of $H^N$ at sites $N$ and $N+1$, but is
also identified with the maximally isotropic planes of the quadratic form
$\Qq^z_N$. These planes are in turn in the domain of the stereographic
projection and, up to a sign, their stereographic projection is precisely the
Green matrix for $H^N$ with that particular boundary condition. A zero measure
set of those boundary conditions contains an anti-Dirichlet part and this set
is not reached in Theorem~\ref{theo-Weyl}(i). Similarly, for semi-infinite
Jacobi matrices, the Weyl limit surface will encode precisely the possible
boundary conditions at infinity. Some part of these boundary conditions are
automatically fixed by the condition of square-integrability, but some other
part may still have to be fixed in order to define a maximal operator.

\section{The Weyl limit disc of a semi-infinite Jacobi matrix}
\label{sec-semiinfinite}

A recursion relation as \eqref{eq-jacobi}, but with $N=\infty$, is called a
semi-infinite Jacobi matrix with matrix entries. The right boundary condition
$Z$ does not intervene (it reappears in the study of extensions below), and the
left boundary condition $\hat{Z}=\hat{Z}^*$ is incorporated in $V_1$ as in
\eqref{eq-matrix}, so that we effectively work with Dirichlet boundary
conditions. Resuming, we consider a formal semi-finite matrix $H$ acting on
$\phi=(\phi_n)_{n\geq 1}\in (\CC^L)^\NN$ by
\begin{equation}
\label{eq-infinitejacobi} (H\,\phi)_n \;=\; T_{n+1}\phi_{n+1}\,+\,V_n\phi_n
\,+\,\delta_{n\neq 1}\,T_{n}^*\phi_{n-1} \;, \qquad n\geq 1\;,
\end{equation}
where $(T_n)_{n\geq 2}$ are invertible and $(V_n)_{n\geq 1}$ self-adjoint
$L\times L$ matrices, and $\delta_{n\neq 1}$ is a Kronecker delta equal to $1$
unless $n=1$, for which it is equal to $0$. Below $H$ will be studied as an
operator on the Hilbert space $\Hh=\ell^2(\NN,\CC^L)$. Associated to $H$ are
also the objects of the previous sections for all $N$. In particular, one has
the matricial solutions $\psi^{\mbox{\rm\tiny D},z}=(\psi^{\mbox{\rm\tiny
D},z}_n)_{n\geq 1}$ and $\psi^{\mbox{\rm\tiny A},z}=(\psi^{\mbox{\rm\tiny
A},z}_n)_{n\geq 1}$, the radial operators $(R^z_N)_{N\geq 1}$, the Weyl discs
$\WM_N^z$, etc.. We shall consider $\psi^{\mbox{\rm\tiny D},z}$ and
$\psi^{\mbox{\rm\tiny A},z}$ as operators from $\CC^L$ to $(\CC^L)^\NN$, {\it
e.g.} $(\psi^{\mbox{\rm\tiny D},z}_nv)_{n\geq 1}\in (\CC^L)^\NN$ for $v\in
\CC^L$. Let us note that $\psi^{\mbox{\rm\tiny A},z}$ does not contain
$\psi^{\mbox{\rm\tiny A},z}_0=\one$ and hence
$$
(H-z\one)\psi^{\mbox{\rm\tiny A},z}\;=\;-\,\pi_1\;,
$$
where $\pi_1:\CC^L\to (\CC^L)^\NN$ denotes the embedding to the first site,
namely $\pi_1 v$ for $v\in \CC^L$ is the vector in $(\CC^L)^\NN$ for which only
the first $L$ entries do not vanish and are given by those of $\phi$.

\begin{defini}
\label{def-limitdisc} Let $\Im m(z)\neq 0$. The closed Weyl limit disc is
defined by
$$
\overline{\WM^z}\;=\; \bigcap_{N\geq 1}\;\overline{\WM^z_N} \;.
$$
\end{defini}

The purpose of this section is to provide several geometric interpretations and
explicit formulas for the Weyl limit disc, and in particular, examine its
maximal boundary, the Weyl limit surface. Helpful will be the following
objects.

\begin{proposi}
\label{prop-limitrad} Let $\Im m(z)\neq 0$. The limits
$$
\Rr^z\;=\;\lim_{N\to\infty}\;\left[\,2\,\Im
m(z)\,\langle\psi^z\,|\,\psi^z\rangle_N\,\right]^{-1}\;,\qquad
R^z\;=\;\lim_{N\to\infty}\;R_N^z\;,\qquad S^z\;=\; \lim_{N\to\infty}\;S_N^z\;,
$$
exist and define matrix-valued functions. Moreover, $\Rr^z$,
$(-\Rr^{\overline{z}})$, $R^z$ and $(-R^{\overline{z}})$ are positive
semi-definite for $\Im m(z)> 0$. Furthermore, $(S^z)^*=S^{\overline{z}}$.
\end{proposi}

\vspace{.2cm}

\noindent {\bf Proof.} Clearly $(\langle\psi^z\,|\,\psi^z\rangle_N)^{-1}$ is a
decreasing sequence of positive matrices. Hence the limit exists. Due to
Proposition~\ref{prop-basicformulas}(iii), $R^z_N$ and $-R^{\overline{z}}_N$
are decreasing sequences of positive definite matrices, so that they converge
to a positive semi-definite matrices $R^z$ and $-R^{\overline{z}}$ if  $\Im
m(z)> 0$. The fact that the center $S^z_N$ of the Weyl discs also converges
follows from Theorem~\ref{theo-Weyl}(iii). \hfill $\Box$

\vspace{.2cm}

Let us stress that the definition of $\Rr^z$ and $R^z$ is analogous by
Proposition~\ref{prop-basicformulas}(ii), one being defined with $\psi^z$ and
the other with $\psi^{\mbox{\rm\tiny D},z}$. By Proposition~\ref{prop-limitrad}
the following orthogonal projections in $\CC^L$ are well-defined:
$$
P^z_0\;=\;\mbox{\rm projection on Ker}(R^z)\;, \qquad P^z_+\;=\;\mbox{\rm
projection on Ker}(R^z)^{\perp}\;=\one\,-\,P^z_0\;.
$$
Let us also set $n_z=\dim(P^z_+)$. In Proposition~\ref{prop-deficiency} below,
these integers will be identified with the deficiency indices of $H$. Hence
they only depend on whether $z$ is in the upper or lower half-plane. The same
holds for the dimension of the kernel of $\Rr^z$ which will also play a r\^ole
shortly.


\begin{proposi}
\label{prop-limitdisc} The Weyl limit disc $\overline{\WM^z}$ is closed. One
has $\overline{\WM^z}=\bigcap_{N\geq 1}\;\WM^z_N$ and
\begin{eqnarray}
\overline{\WM^z} \label{eq-limitdics1} & = & \left\{\,S^z+(R^z)^{\frac{1}{2}} W
(-R^{\overline{z}})^{\frac{1}{2}} \;\left|\;
W=P_+^zWP_+^{\overline{z}}\in\overline{\DM_L}\;\right.\right\} \\
& = & \left\{\, \, G\in\, \mbox{\rm Mat}(L,\CC)\;\left|
\;\,\lim_{N\to\infty}\,\mbox{\rm sgn} (\Im
m(z))\;\Phi_G^*\Qq^z_N\Phi_G\,\leq\,\nul \;\right.\right\}
\label{eq-limitdics2} \\
& = & \left\{\, \, G\in\, \mbox{\rm Mat}(L,\CC)\;\left| \;\,2\,|\Im
m(z)|\;\sum_{n=1}^\infty\left|\psi^{\mbox{\rm\tiny
D},{z}}_nG-\psi^{\mbox{\rm\tiny
A},{z}}_n\right|^2\;\leq\;\imath\,(G^*-G)\;\right.\right\}\;.\label{eq-limitdics3}
\end{eqnarray}
\end{proposi}

\vspace{.2cm}

\noindent {\bf Proof.} The equality $\cap_{N\geq 1}\WM^z_N=\cap_{N\geq
1}\overline{\WM^z_N}$ follows from the fact that the Weyl discs are strictly
nested by Theorem~\ref{theo-Weyl}(iii). The first equality follows directly
from the definition of $\WM^z_N$ and $P^z_+$ as well as
Proposition~\ref{prop-limitrad}. The second equality follows by taking the
limit of \eqref{eq-Weylrewrite2} and recalling that $\sigma^z\Qq^z_N$ is
non-decreasing in $N$. The last equality is obtained either by rewriting
\eqref{eq-limitdics2} or by taking the limit of \eqref{eq-Weylsurfacefrom3}.
\hfill $\Box$

\vspace{.2cm}

For the construction of extensions, a central r\^ole will be played by the
maximal boundary of the Weyl limit discs.

\begin{defini}
\label{def-limitsurface} The Weyl limit surface for $\Im m(z)\neq 0$ is defined
by
\begin{equation}
\label{eq-limitsurface}
\partial_\mx\WM^z \;=\;
\left\{\, \, G\in\, \mbox{\rm Mat}(L,\CC)\;\left|
\;\,\lim_{N\to\infty}\,\Phi_G^*\Qq^z_N\Phi_G\,=\,\nul \;\right.\right\}\;.
\end{equation}
If $R^z=0$ and $R^{\overline{z}}=0$, then $\overline{\WM^z}=\partial_\mx\WM^z$
and $\overline{\WM^{\overline{z}}}=\partial_\mx\WM^{\overline{z}}$ consist of a
single point and one says that $H$ is in the limit point case. If $R^z>\nul$
and $(-R^{\overline{z}})>\nul$, one calls $H$ completely indeterminate {\rm
(}following Krein {\rm \cite{Kre})}.
\end{defini}

In Proposition~\ref{prop-Weylsurfacedescription} below it will be shown that
the Weyl limit surface can also be expressed in purely geometric terms, just as
in the finite volume case of Definition~\ref{def-Weyldisc}. Let us note that
according to Definition~\ref{def-limitsurface} the closed Weyl limit disc can
consist of only one point, without $H$ being limit point. More precisely, if
$R^{\overline{z}}=\nul$, then $\partial_\mx\WM^z=\{S^z\}$, but if, moreover,
$R^{{z}}\neq\nul$, then $\partial_\mx\WM^z\neq\{S^{\overline{z}}\}$. Clearly,
it is possible to rewrite the definition \eqref{eq-limitsurface} in the form
\eqref{eq-limitdics3}, albeit with an equality. Moreover, the representation
\eqref{eq-limitdics1} shows
\begin{equation}\label{eq-Weyllimitrep}
\partial_\mx\WM^z \;=\;
\left\{\,S^z+(R^z)^{\frac{1}{2}} W (-R^{\overline{z}})^{\frac{1}{2}} \;\left|\;
W=P^z_+WP^{\overline{z}}_+ \;\;\mbox{\rm partial isometry, injective on
}P^{\overline{z}}_+\CC^L\;\right.\right\}\;.
\end{equation}

\vspace{.2cm}

In \eqref{eq-limitdics2} and \eqref{eq-limitsurface} appears the limit of the
quadratic forms $\Qq^z_N$. If $\Im m(z)>0$, then the sequence $(\Qq_N^z)_{N\geq
1}$ is increasing by Proposition~\ref{prop-Qproperties}(iv). The eigenvalues of
$\Qq^z_N$ thus increase with $N$ and may diverge to $+\infty$. Moreover, for
any $v\in\CC^{2L}$ the sequence $v^*\Qq^z_Nv$ is increasing and converges
therefore either to a finite number or $+\infty$. Due to the Cauchy-Schwarz
inequality, the set $\{v\in\CC^{2L}\,|\,\lim_{N\to\infty}v^*\Qq^z_Nv<\infty\}$
is linear subspace in $\CC^{2L}$. Let $\Pp^z_\infty$ denote the orthogonal
projection in $\CC^{2L}$ on its orthogonal complement. Then
\begin{equation}
\label{eq-dicho}
\lim_{N\to\infty} \,w^*\Qq^z_Nw\;\;\mbox{exists} \quad
\Longleftrightarrow \quad \Pp^z_{\infty} w\,=\,0\;.
\end{equation}
Hence $\Qq^z=\lim_{N\to\infty} (\one-\Pp^z_\infty)\Qq^z_N(\one-\Pp^z_\infty)$
defines a quadratic form on $(\one-\Pp^z_\infty)\CC^{2L}$. Let $\Pp^z_-$ and
$\Pp^z_+$ denote the spectral projections on the negative and positive
eigenvalues of $\Qq^z$ and $\Pp^z_0$ the projections on its kernel. Hence
$\Pp^z_-+\Pp^z_0+\Pp^z_++\Pp^z_\infty=\one$. Similarly, still for $\Im m(z)>0$,
$\Qq^{\overline{z}}_N$ is decreasing and defines in the limit projections
$\Pp^{\overline{z}}_{-\infty}$, $\Pp^{\overline{z}}_{-}$,
$\Pp^{\overline{z}}_0$ and $\Pp^{\overline{z}}_{+}$, as well as a self-adjoint
operator $\Qq^{\overline{z}}$ on $(\one-\Pp^{\overline{z}}_{-\infty})\CC^{2L}$.
Somewhat abusing notations, we will write $\Pp^{\overline{z}}_{\infty}$ instead
of $\Pp^{\overline{z}}_{-\infty}$. Using these definitions and facts, it is
possible to rewrite \eqref{eq-limitsurface} as
\begin{equation}
\label{eq-Weyllimitsurface0}
\partial_\mx\WM^z \;=\;
\left\{\, \, G\in\, \mbox{\rm Mat}(L,\CC)\;\left|
\;\,\Pp^z_\infty\Phi_G=\nul\;,\;\; \,\Phi_G^*\Qq^z\Phi_G\,=\,\nul
\;\right.\right\}\;.
\end{equation}
Similarly, $\overline{\WM^z}$ is given by the condition
$\Phi_G^*\Qq^z\Phi_G\leq \nul$. Moreover, one can show that $\Pp^z_\infty$ is
the projection on the kernel of $\Rr^z$:

\vspace{.2cm}

\begin{proposi}
\label{prop-eliminate} Let $w\in\CC^{2L}$. Then
$$
\Pp^z_{\infty} w\,=\,0 \quad\Longleftrightarrow \quad \lim_{N\to\infty}\;
\langle \psi^zw\,|\,\psi^zw\rangle_N\,<\,\infty \quad \Longleftrightarrow \quad
w\in \mbox{\rm Ker}(\Rr^{z})^\perp \;.
$$
If $G\in\overline{\WM^z}$, then $\Pp_{\infty}^z\,\Phi_G=0$.
\end{proposi}

\noindent {\bf Proof.} By \eqref{eq-dicho}, one has $\Pp_{\infty}^z\,w=0$ if
and only if $\lim_{N\to\infty}w^*\Qq^z_Nw<\infty$ (or $>-\infty$ if $\Im
m(z)<0$). But by Proposition~\ref{prop-basicformulas}(ii),
\begin{equation}
\label{eq-formeq} \lim_{N\to\infty} \;w^*\Qq^z_Nw\;=\;\frac{1}{\imath}\;w^*\Jj
w\,+\, 2\,\Im m(z)\;\lim_{N\to\infty}\; \langle \psi^zw\,|\,\psi^zw\rangle_N\;,
\end{equation}
implying the first equivalence. The second equivalence follows from the
definition of $\Rr^z$ in Proposition~\ref{prop-limitrad}. The last statement
follows from identity \eqref{eq-limitdics3} in
Proposition~\ref{prop-limitdisc}. \hfill $\Box$

\vspace{.2cm}

There are matrices $G$ such that $\Pp_{\infty}^z\Phi_G=0$ which are not in
$\overline{\WM^z}$, that is $\Phi_G^*\Qq^z\Phi_G\leq 0$ does not hold. On the
other hand, $\Phi_G^*\Qq^z\Phi_G$ is then always bounded above by
$C\,\Phi_G^*\Phi_G$ where $C$ is the maximum of the spectrum of $\Qq^z$.
Particularly important for a good understanding of the deficiency spaces will
be the following results.

\vspace{.2cm}

\begin{proposi}
\label{prop-dimensions} Let $\Im m(z)>0$.

\vspace{.1cm}

\noindent {\rm (i)}
$L=\dim(\Pp^z_-)+\dim(\Pp^z_0)=\dim(\Pp^z_+)+\dim(\Pp^z_\infty)$

\vspace{.1cm}

\noindent {\rm (ii)} $L=\dim(\Pp^{\overline{z}}_-)+\dim(\Pp^{\overline{z}}_0)=
\dim(\Pp^{\overline{z}}_+)+\dim(\Pp^{\overline{z}}_{-\infty})$

\vspace{.1cm}

\noindent {\rm (iii)} $\Pp^z_\infty=\Jj\Pp^{\overline{z}}_0\Jj^*$,
$\Pp^z_0=\Jj\Pp^{\overline{z}}_{\infty}\Jj^*$ and
$\Pp^z_\pm=\Jj\Pp^{\overline{z}}_\pm\Jj^*$

\vspace{.1cm}

\noindent {\rm (iv)}
$\dim(\Pp^z_\infty)=\dim(P^{{z}}_0)=\dim(\Pp^{\overline{z}}_0)$ and
$\dim(\Pp^z_+)=\dim(P^z_+)=\dim(\Pp^{\overline{z}}_+)$

\vspace{.1cm}

\noindent {\rm (v)}
$\dim(\Pp^{\overline{z}}_{\infty})=\dim(P^{\overline{z}}_0)=\dim(\Pp^z_0)$ and
$\dim(\Pp^{\overline{z}}_-)=\dim(P^{\overline{z}}_+)=\dim(\Pp^z_-)$

\end{proposi}

\vspace{.2cm}

\noindent {\bf Proof.} Item (i) and (ii) follow from Sylvester's law given in
Proposition~\ref{prop-Qproperties}(v). The equalities in (iii) are a direct
consequence of the identity $(\Qq^z_N)^{-1}=\Jj\Qq^{\overline{z}}_N\Jj^*$
proven in Proposition~\ref{prop-Qproperties}(iii). Now let us turn to the proof
of (iv). Let $v=P^z_0v\in\ker(R^z)\subset \CC^L$. Then it follows from
Proposition~\ref{prop-Qproperties}(ii) and
Proposition~\ref{prop-basicformulas}(i) that
$$
\lim_{N\to\infty}\; \left(\begin{array}{c}
v \\
0
\end{array}
\right)^* \Qq^z_N \left(
\begin{array}{c}
v \\
0
\end{array}
\right) \;=\; \lim_{N\to\infty} v^*(R^z_N)^{-1}v\;=\;\infty\;.
$$
Hence, by the minimax principle, $\dim(\Pp^z_\infty)\geq\dim(P^{{z}}_0)$.
Arguing similarly with $v=P^z_+v\in\ker(R^z)^{\perp}$ shows
$\dim(\Pp^z_+)\geq\dim(P^z_+)$. As $\dim(P^{{z}}_+)+\dim(P^{{z}}_0)=L$, the
second equality in item (i) proves that the two inequalities actually both have
to be equalities. The remaining equalities of (iv) follow from (iii). Finally
(v) is proven similarly, noting the different sign in
Proposition~\ref{prop-Qproperties}(ii), however. \hfill $\Box$

\vspace{.2cm}

\begin{proposi}
\label{prop-maxiso} The maximally isotropic subspaces of the quadratic from
$\Qq^z$ on $(\one-\Pp^z_\infty)\CC^{2L}$ are of dimension
$$
\dim(\Pp_0^z)\,+\,\min\{\dim(\Pp^z_+),\dim(\Pp^z_-)\}\;
=\;\dim(\Pp_0^z)\,+\,\min\{n_z,n_{\overline{z}}\}\;.
$$
{\rm (}This number is also called the Witt index of $\Qq^z$.{\rm )} When $\Im
m(z)>0$, this dimension is $L$ if and only if $n_z\leq n_{\overline{z}}$ {\rm
(}and $n_z\geq n_{\overline{z}}$ whenever $\Im m(z)<0${\rm )}.
\end{proposi}

\noindent {\bf Proof.} Every maximally isotropic subspace of $\Qq^z$ contains
the eigenspace of $\Qq^z$ corresponding to the eigenvalue $0$ which is given by
$\Pp^z_0\CC^{2L}$. For the non-degenerate form on $(\Pp^z_++\Pp^z_-)\CC^{2L}$
the dimension of the maximally isotropic subspace is given by the minimum of
the dimensions of the subspace with positive and negative eigenvalues. All the
remaining statements follow from Proposition~\ref{prop-dimensions}. \hfill
$\Box$

\vspace{.2cm}

We shall say that a plane described by a $2L\times n$ matrix $\Phi$ is
isotropic for $\Qq^z$ if and only if $\Pp^z_\infty\Phi=\nul$ and
$\Phi^*\Qq^z\Phi=\nul$. Of course, for every such isotropic plane one has
rank$(\Phi)\leq L$. Now it can be shown that all isotropic planes of $\Qq^z$
are of the type appearing in \eqref{eq-Weyllimitsurface0}:

\vspace{.2cm}

\begin{proposi}
\label{prop-Weylsurfacedescription} Let $\Im m(z)>0$ and $n_z\leq
n_{\overline{z}}$. Then
\begin{equation}
\label{eq-Weyllimitsurface} \partial_\mx\WM^z\;=\; -\,\pi\left(\,\left\{\,
[\Phi]_\sim\in\GM_L
 \;\left|\;
\Phi\;\;\mbox{\rm isotropic for }\Qq^z
 \;\right.\right\}\;\right)\,.
\end{equation}
For $\Im m(z)<0$ the same representation is valid whenever $n_z\geq
n_{\overline{z}}$.
\end{proposi}

\vspace{.2cm}

\noindent {\bf Proof.} Due to \eqref{eq-Weyllimitsurface0} and
$\pi([\Phi_G]_\sim)=-G$, it is clear that the Weyl limit surface
$\partial_\mx\WM_N^z$ is contained in the set on the r.h.s. of
\eqref{eq-Weyllimitsurface}. For the converse, let us first note that the
isotropic planes of $\Qq^z$ are indeed $L$-dimensional by
Proposition~\ref{prop-maxiso} (because $n_z\leq n_{\overline{z}}$). We have to
show that each such subspace $[\Phi]_\sim$ is represented by some $\Phi_G$ with
$G\in\UM_L$. By Lemma~\ref{lem-Wronskinv}, it is sufficient to show that
$\Ww(\Phi,\Phi)<\nul$. Because $\Pp^z_\infty\Phi=\nul$, \eqref{eq-formeq} and
Proposition~\ref{prop-limitrad} one has
$$
\Ww(\Phi,\Phi)\;=\;-\,\Phi^*\,(\one-\Pp^z_\infty)\, (\Rr^z)^{-1}\,
(\one-\Pp^z_\infty)\,\Phi\;.
$$
But $(\one-\Pp^z_\infty)\Phi=\Phi$ has rank $L$ and lies in the orthogonal
complement of the kernel of $\Rr^z$ on which $(\Rr^z)^{-1}$ is strictly
positive. Hence the r.h.s. negative. \hfill $\Box$

\vspace{.2cm}

It will be extremely useful to introduce solutions which are normalized on the
square integrable subspaces and vanish on the others:
\begin{equation}
\label{eq-normalizeddef} \tilde{\psi}^{z}\;=\; \psi^{z}\, \bigl(
(z-\overline{z})\Rr^z\bigr)^{\frac{1}{2}} \;, \qquad
\tilde{\psi}^{\mbox{\rm\tiny D},z}\;=\; \psi^{\mbox{\rm\tiny D},z}\, \bigl(
(z-\overline{z})R^z\bigr)^{\frac{1}{2}} \;, \qquad \Im m(z)\neq 0 \;.
\end{equation}
For all $w\in\CC^{2L}$ and $v\in\CC^L$ one then has $\tilde{\psi}^zw\in\Hh$ and
$\tilde{\psi}^{\mbox{\rm\tiny D},z}v\in\Hh$. Moreover,
$H\tilde{\psi}^zw=z\tilde{\psi}^zw$ and $H\tilde{\psi}^{\mbox{\rm\tiny
D},z}v=z\tilde{\psi}^{\mbox{\rm\tiny D},z}v$. Furthermore $\tilde{\psi}^z$ and
$\tilde{\psi}^{\mbox{\rm\tiny D},z}$ are partial isometries. More precisely,
$(\tilde{\psi}^{z})^*\tilde{\psi}^{z}=\one-\Pp^z_\infty$ and, if $\Pi^z$
denotes the orthogonal projection in $\Hh$ onto $N_z$,
\begin{equation}
\label{eq-defiso} (\tilde{\psi}^{\mbox{\rm\tiny
D},z})^*\tilde{\psi}^{\mbox{\rm\tiny D},z}\;=\;P^z_+\;,\qquad
\tilde{\psi}^{\mbox{\rm\tiny D},z}(\tilde{\psi}^{\mbox{\rm\tiny
D},z})^*\;=\;\Pi^z\;.
\end{equation}
Let us also introduce $\Pi^{z,\overline{z}}$ as the projection on $N_z\dotplus
N_{\overline{z}}$. Note that $\Pi^{z,\overline{z}}$ is not equal to
$\Pi^{\overline{z}}+\Pi^{z}$ because $N_z$ and $N_{\overline{z}}$ are not
orthogonal, and it is also not equal to the projection
$\tilde{\psi}^{z}(\tilde{\psi}^{z})^*$. The latter actually dominates
$\Pi^{z,\overline{z}}$. It is, nevertheless, possible to define a partial
isometry from $\Pi^{z,\overline{z}}\Hh$ to $(\Pp^z_++\Pp^z_-)\CC^{2L}$ as
follows:
$$
\hat{\psi}^z\;=\; \psi^{z}\, \bigl(
(z-\overline{z})\hat{\Rr}^z\bigr)^{\frac{1}{2}} \;, \qquad
\hat{\Rr}^z\;=\;\lim_{N\to\infty}\;\left[\,2\,\Im
m(z)\,\langle\psi^z\,|\,(\Pp^z_++\Pp^z_-)\,
|\,\psi^z\rangle_N\,\right]^{-1}\;(\Pp^z_++\Pp^z_-) \;.
$$
Then $\hat{\psi}^z(\hat{\psi}^z)^*=\Pi^{z,\overline{z}}$ and
$(\hat{\psi}^z)^*\hat{\psi}^z=\Pp_+^z+\Pp_-^z$. Using
Proposition~\ref{prop-dimensions}(iii), the latter projection can also be
expressed in terms of $\tilde{\psi}^z$:
$$
\Pp_+^z+\Pp_-^z\;=\;(\tilde{\psi}^z)^*\tilde{\psi}^z
\,+\,\Jj\,(\tilde{\psi}^{\overline{z}})^* \tilde{\psi}^{\overline{z}} \,\Jj^*
\,-\,\one\;.
$$
As an application of these notations, let us show how $R^z$ and $S^z$ appear in
formulas for limit Wronskians of the square integrable solutions. All these
formulas will be used for the characterization of the domain of the adjoint of
$H$ in the next section.

\vspace{.2cm}

\begin{proposi}
\label{prop-limitWronskianformulas} Let $\Im m(z)\neq 0$.

\vspace{.1cm}

\noindent {\rm (i)} $\lim_{N\to\infty} \Ww\bigl(\tilde{\Psi}_N^{\mbox{\rm\tiny
D},z}, \tilde{\Psi}_N^{\mbox{\rm\tiny
D},\overline{z}}\bigr)=\nul=\lim_{N\to\infty} \Ww\bigl({\Psi}_N^{\mbox{\rm\tiny
D},z}, \tilde{\Psi}_N^{\mbox{\rm\tiny D},\overline{z}}\bigr)$

\vspace{.1cm}

\noindent {\rm (ii)} $\lim_{N\to\infty} \Ww\bigl(\tilde{\Psi}_N^{\mbox{\rm\tiny
D},z}, \tilde{\Psi}_N^{\mbox{\rm\tiny D},{z}}\bigr)=(z-\overline{z})P^z_+\;$
and $\;\lim_{N\to\infty}\Ww\bigl({\Psi}_N^{\mbox{\rm\tiny D},z},
\tilde{\Psi}_N^{\mbox{\rm\tiny
D},{z}}\bigr)=(z-\overline{z})\bigl((z-\overline{z})R^z\bigr)^{-\frac{1}{2}}P^z_+$

\vspace{.1cm}

\noindent {\rm (iii)} $\lim_{N\to\infty} \Ww\bigl({\Psi}_N^{\mbox{\rm\tiny
A},z}, \tilde{\Psi}_N^{\mbox{\rm\tiny D},{z}}\bigr)=
(z-\overline{z})S^{\overline{z}}\bigl((z-\overline{z})R^z\bigr)^{-\frac{1}{2}}P^z_+$

\vspace{.1cm}

\noindent {\rm (iv)} $\lim_{N\to\infty} \Ww\bigl({\Psi}_N^{\mbox{\rm\tiny
A},z}, \tilde{\Psi}_N^{\mbox{\rm\tiny D},\overline{z}}\bigr)=
-\,\imath\,\bigl((\overline{z}-z)R^{\overline{z}}\bigr)^{\frac{1}{2}}P^{\overline{z}}_+$

\end{proposi}

\vspace{.2cm}

\noindent {\bf Proof.} All these formulas result from
Propositions~\ref{prop-basicWronskian} and \ref{prop-basicformulas}. For (i),
one has $\Ww( \tilde{\Psi}_N^{\mbox{\rm\tiny D},{z}},
\tilde{\Psi}_N^{\mbox{\rm\tiny D},\overline{z}})=\nul$ even for finite $N$. For
(ii),
$$
\lim_{N\to\infty}\, \Ww\bigl(\tilde{\Psi}_N^{\mbox{\rm\tiny D},z},
\tilde{\Psi}_N^{\mbox{\rm\tiny D},{z}}\bigr)\;=\;
\lim_{N\to\infty}\,\bigl((z-\overline{z})R^z\bigr)^{\frac{1}{2}}
\,(z-\overline{z})\,\langle{\psi}^{\mbox{\rm\tiny
D},{z}}|{\psi}^{\mbox{\rm\tiny D},{z}}\rangle_N
\,\bigl((z-\overline{z})R^z\bigr)^{\frac{1}{2}} \;=\;
(z-\overline{z})\,P^z_+\;.
$$
Similarly,
$$
\lim_{N\to\infty}\, \Ww\bigl({\Psi}_N^{\mbox{\rm\tiny A},z},
\tilde{\Psi}_N^{\mbox{\rm\tiny D},{z}}\bigr)\,=\, \lim_{N\to\infty} \left(
\begin{array}{c}
\nul \\
\one
\end{array}
\right)^* \Qq^z_N \left(
\begin{array}{c}
\one \\
\nul
\end{array}
\right) \bigl((z-\overline{z})R^z\bigr)^{\frac{1}{2}} \,=\, \lim_{N\to\infty}
(S^z_N)^*(R^z_N)^{-1}\bigl((z-\overline{z})R^z\bigr)^{\frac{1}{2}}\;,
$$
from which (iii) follows. The proof of (iv) is the similar. \hfill $\Box$

\vspace{.2cm}

\section{The adjoint of a semi-infinite Jacobi matrix} \label{sec-semiadjoint}

The aim of this section is to examine the operators that the prescription $H$
given by \eqref{eq-infinitejacobi} defines on the Hilbert space
$\Hh=\ell^2(\NN,\CC^L)$. All operators constructed from $H$ will be denoted by
bold face letters. The mininal operator $\HH^0$ is defined on $\Dd({\bf
H}^0)=\{\phi\in\Hh\,|\,\phi_n=0\mbox{ except for finitely many }n\}$, by
setting $\HH^0\phi=H\phi$. Because ${\bf H}^0$ is symmetric, it is closable.
Its closure will be denoted by $\HH$ with domain $\Dd(\HH)$. By definition, the
domain of its adjoint is
$$
\Dd(\HH^*)\;=\; \{\, \phi\in\Hh\,|\,\phi'\in\Dd(\HH)\mapsto \langle
\phi|\HH\phi'\rangle\mbox{ bounded functional}\,\}\;.
$$
Moreover, by general principle $\Dd(\HH^*)=\Dd((\HH^0)^*)$. But for
$\phi,\phi'\in\Dd((\HH^0)^*)$, one has $\langle \phi|\HH^0\phi'\rangle=\langle
H\phi|\phi'\rangle$ because the sum in the scalar product is finite. As
$\Dd(\HH^0)$ is dense in $\Hh$, the Riesz theorem therefore implies
$$
\Dd(\HH^*)\;=\; \{\, \phi\in\Hh\,|\,H\phi\in\Hh\;\}\;,\qquad
\HH^*\phi\,=\,H\phi\;.
$$
By the above, one has $\tilde{\psi}^zw,\tilde{\psi}^{\mbox{\rm\tiny
D},z}v\in\Dd(\HH^*)$ for all $w\in\CC^{2L}$ and $v\in\CC^L$. Next recall that
the deficiency subspaces of $\HH$ are defined by
\begin{equation}
\label{eq-Ndef} N_z\;=\; \mbox{\rm Ran}(\HH-\overline{z}\,\one)^\perp\;=\;
\mbox{\rm Ker}(\HH^*-{z}\,\one) \;.
\end{equation}
and the deficiency indices by $\dim(N_z)$. It is well-known that the deficiency
index $\dim(N_z)$ only depends on whether $z$ is in the upper or lower
half-plane. For their analysis, let us begin by recalling the following general
fact.

\begin{proposi}
\label{prop-domain}  For $\Im m(z)\neq 0$, one has $\Dd(\HH^*)=\Dd(\HH)\dotplus
N_z\dotplus N_{\overline{z}}$.
\end{proposi}

\vspace{.2cm}

\noindent {\bf Proof.} Let $\psi\in\Dd(\HH^*)$. Then
$(\HH^*-z\,\one)\psi\in\Hh$. As $\Hh=\,$Ran$(\HH-z\,\one)\oplus
N_{\overline{z}}$ and $\Im m(z)\neq 0$, there exist $\varphi\in \Dd(\HH)$ and
$\phi\in N_{\overline{z}}$ such that
$$
(\HH^*-z\,\one)\psi\;=\;(\HH-z\,\one)\varphi + (z-\overline{z})\phi\;.
$$
But $\phi\in N_{\overline{z}}$ implies $\HH^*\phi=\overline{z}\phi$. Because
$\HH^*$ is an extension of $\HH$, it follows that
$(\HH^*-z\,\one)(\psi-\varphi-\phi)=0$, that is $(\psi-\varphi-\phi)\in N_z$.
Therefore $\Dd(\HH^*)=\Dd(\HH)+N_z+N_{\overline{z}}$. In order to show that the
decomposition is direct, let $\psi\in N_z$ be decomposed as $\psi=\varphi +
\phi$ with $\varphi\in\Dd(\HH)$ and $\phi\in N_{\overline{z}}$. Then
$0=(\HH^*-z\one)(\varphi+\phi)=(\HH-z\one)\varphi+(\overline{z}-z)\phi$. But
$(\HH-z\one)\varphi\in\mbox{\rm Ran}(\HH-{z}\,\one)$ and $\phi\in\mbox{\rm
Ran}(\HH-{z}\,\one)^\perp$, so that $(\HH-z\one)\varphi=0$ and $\phi=0$. As
$\HH$ is symmetric, $\varphi=0$ and thus $\psi=0$. Similarly, $\psi\in
N_{\overline{z}}$ vanishes if it is given by a linear combination of elements
in $\Dd(\HH)$ and $N_z$. Finally, let $\psi\in\Dd(\HH)$ be given by
$\psi=\phi+\varphi$ with $\phi\in N_{\overline{z}}$ and $\varphi\in N_z$. Then
$0=\langle \phi|(\HH-z\one)\psi\rangle=\langle
\phi|(\HH-z\one)(\phi+\varphi)\rangle = (\overline{z}-z)\|\phi\|^2$ so that
$\phi=0$ and similarly $\varphi=0$. (Alternatively, one can verify that the
decomposition is orthogonal w.r.t. the graph scalar product, see \cite{Sim}.)
\hfill $\Box$

\vspace{.2cm}

The aim of the following results is to calculate the deficiency spaces and
hence the domain of the adjoint explicitly, as well as the action of the
adjoint on the Dirichlet and anti-Dirichlet solutions.

\vspace{.2cm}

\begin{proposi}
\label{prop-deficiency} Let $\Im m(z)\neq 0$, $v\in\CC^L$ and $w\in\CC^{2L}$.

\vspace{.1cm}

\noindent {\rm (i)} $\psi^{z}w\in\Hh$ if and only if $\Pp^z_{\infty} w=0$.

\vspace{.1cm}

\noindent {\rm (ii)}  $\psi^{\mbox{\rm\tiny D},z}v\in\Hh$ if and only if
$P^z_0v=0$.

\vspace{.1cm}

\noindent {\rm (iii)} $\psi^{\mbox{\rm\tiny A},z}v\in\Hh$ if and only if $v\in
(\overline{\WM^z})^{-1}(\mbox{\rm Ker}(R^z)^\perp)$.

\vspace{.1cm}

\noindent {\rm (iv)} One has
$$
N_z\;=\;\bigl\{\,\left.\psi^{\mbox{\rm\tiny D},z}v\,=\,(\psi^{\mbox{\rm\tiny
D},z}_nv)_{n\geq 1}\in\Hh\;\right|\;v\in \mbox{\rm Ker}(R^z)^\perp \bigr\} \;,
\qquad n_z\;=\;\dim(N_z)\;.
$$

\noindent {\rm (v)} The only solutions $\phi\in\Hh$ of $(\HH^*-z\,\one)\phi=0$
are $\psi^{\mbox{\rm\tiny D},z}v$ with $v\in\mbox{\rm
Ker}(R^z)^\perp=P^z_+\CC^L$.

\vspace{.1cm}

\noindent {\rm (vi)} The only solutions $\phi\in\Hh$ of
$(\HH^*-z\,\one)\phi=-\,\pi_1\pi_1^*\phi$ are $\psi^{\mbox{\rm\tiny A},z}v$
with $v\in (\overline{\WM^z})^{-1}(\mbox{\rm Ker}(R^z)^\perp)$.

\vspace{.1cm}

$\;$  In particular, $\psi^{\mbox{\rm\tiny A},z}v\in\Dd(\HH^*)$ for $v\in
(\overline{\WM^z})^{-1}(\mbox{\rm Ker}(R^z)^\perp)$.

\vspace{.1cm}

\noindent {\rm (vii)} For $G\in \overline{\WM^z}$, one has
$(\psi^{\mbox{\rm\tiny D},z}G-\psi^{\mbox{\rm\tiny A},z})v\in\Hh$ for all
$v\in\CC^L$ and
\begin{equation}
\label{eq-actDir0} (\HH^*-{z}\,\one)(\psi^{\mbox{\rm\tiny
D},z}G-\psi^{\mbox{\rm\tiny A},z})\;=\;\pi_1\;.
\end{equation}

\end{proposi}

\vspace{.2cm}

\noindent {\bf Proof.} (i) is just restating Proposition~\ref{prop-eliminate}
and (ii) is proven similarly.
%

\vspace{.1cm}

(iii) The inequality in \eqref{eq-limitdics3} implies that
$(\psi^{\mbox{\rm\tiny D},z}G-\psi^{\mbox{\rm\tiny A},z})v\in\Hh$ for all
$G\in\overline{\WM^z}$ and $v\in\CC^L$. For $Gv\in\,$Ker$(R^z)^\perp$ one knows
by (i) that $\psi^{\mbox{\rm\tiny D},z}Gv\in\Hh$. Hence for
$v\in(\overline{\WM^z})^{-1}($Ker$(R^z)^\perp)$ one has $\psi^{\mbox{\rm\tiny
A},z}v\in\Hh$. Inversely, if $\psi^{\mbox{\rm\tiny A},z}v\in\Hh$, then for any
$G\in \overline{\WM^z}$ one has $\psi^{\mbox{\rm\tiny
D},z}Gv=(\psi^{\mbox{\rm\tiny D},z}G-\psi^{\mbox{\rm\tiny
A},z})v+\psi^{\mbox{\rm\tiny A},z}v\in\Hh$. By (ii), one concludes
$Gv\in\,$Ker$(R^z)^\perp$.

\vspace{.1cm}

(iv) By definition, $N_z$ is the subspace of $\phi\in \Hh$ such that $\langle
\phi|(\HH-\overline{z}\,\one)\phi'\rangle=0$ for all $\phi'\in\Dd(\HH)$, that
is, such that $\langle (\HH^*-z\,\one)\phi|\phi'\rangle=0$ for all
$\phi'\in\Dd(\HH)$. As $\Dd(\HH)$ is dense, $N_z$ is thus the set of
$\phi\in\Hh$ for which $\HH^*\phi=z\phi$. In particular,
$\HH^*\phi=H\phi\in\Hh$. In conclusion, $N_z$ is the set of (square integrable)
solutions $\phi\in\Hh$ of the equation $H\phi=z\phi$. Due to the arguments in
Section~\ref{sec-finite} and the Dirichlet boundary conditions, all (formal,
i.e. not necessarily square integrable) solutions of $H\phi=z\phi$ are now
given by $\phi=(\psi^{\mbox{\rm\tiny D},z}_nv)_{n\geq 1}$ where $v\in \CC^L$
(anti-Dirichlet solutions of $H\phi=z\phi$ vanish identically). But by (ii),
$\phi\in\Hh$ if and only if $v\in\,$Ker$(R^z)^\perp$.

\vspace{.1cm}

(v) and (vi) It follows from the arguments in Section~\ref{sec-finite} that the
only matricial solutions $\psi=(\psi_n)_{n\geq 1}$ of $(H-z\,\one)\psi=0$ and
$(H-z\,\one)\psi=-\pi_1\pi_1^*\psi$ are $\psi^{\mbox{\rm\tiny D},z}$ and
$\psi^{\mbox{\rm\tiny A},z}$ respectively. Due to (ii) and (iii) they lead
precisely to the stated square integrable solutions in Hilbert space.

\vspace{.1cm}

(vii) The fact that $\psi=(\psi^{\mbox{\rm\tiny A},z}G-\psi^{\mbox{\rm\tiny
A},z})v\in\Hh$ for $G\in\overline{\WM^z}$ follows from \eqref{eq-limitdics3} in
Proposition~\ref{prop-limitdisc}. Then $(H-z\,\one)\psi=\pi_1v$ follows from
Section~\ref{sec-finite}. \hfill $\Box$

\vspace{.2cm}

If the kernel of $\HH^*-{z}\,\one$ is trivial, then \eqref{eq-actDir0} shows
that $G$ is the Green matrix of $\HH^*$. This happens only in the limit point
case. Otherwise, namely either $R^z\neq\nul$ or $R^{\overline{z}}\neq\nul$, one
can show explicitly that the adjoint $\HH^*$ is not symmetric (in particular,
$\HH^*$ is not self-adjoint). For example, for $v\in\mbox{\rm Ker}(R^z)^\perp$,
$v\neq 0$, the proposition shows
$$
\langle \HH^*\psi^{\mbox{\rm\tiny D},z}v\,|\,\psi^{\mbox{\rm\tiny D},z}v\rangle
\;=\;\overline{z}\;\|\psi^{\mbox{\rm\tiny
D},z}v\|^2\;\neq\;z\;\|\psi^{\mbox{\rm\tiny D},z}v\|\;=\; \langle
\psi^{\mbox{\rm\tiny D},z}v\,|\,\HH^*\psi^{\mbox{\rm\tiny D},z}v\rangle \;.
$$
The defect from being self-adjoint can be measured by the so-called limit
Wronskian.

\begin{proposi}
\label{prop-limitwronskian} Let $\phi,\psi\in\Dd(\HH^*)$ and associate
$\Phi_N,\Psi_N$ as in {\rm \eqref{eq-solcorr}}. Then
$$
\langle \phi\,|\,\HH^*\psi\rangle\,-\,\langle \HH^*\phi\,|\,\psi\rangle \;=\;
\imath\;\lim_{N\to\infty}\,\Ww(\Phi_N,\Psi_N) \;.
$$
\end{proposi}

\vspace{.2cm}

\noindent {\bf Proof.} Let $\phi,\psi\in\Dd(\HH^*)$. Then
$$
\langle \phi\,|\,\HH^*\psi\rangle\,-\,\langle \HH^*\phi\,|\,\psi\rangle \;=\;
\lim_{N\to\infty}\,\sum_{n=1}^N\bigl(\,\phi_n^*(H\psi)_n-(H\phi)_n^*\psi_n\,
\bigr)\;.
$$
Replacing the recurrence relation \eqref{eq-infinitejacobi} twice and
telescoping now shows
$$
\langle \phi\,|\,\HH^*\psi\rangle\,-\,\langle \HH^*\phi\,|\,\psi\rangle \; =\;
\lim_{N\to\infty}\,\bigl( \phi_N^*T_{N+1}\psi_{N+1}-
\phi_{N+1}^*T_{N+1}^*\psi_N\bigr)\;.
$$
But the r.h.s. without the limit is precisely $\imath\Ww(\Phi_N,\Psi_N)$.
\hfill $\Box$

\vspace{.2cm}

It may seem that the limit Wronskian on the r.h.s. always vanishes because
$\phi,\psi$ are square-integrable. This is not true because the definition of
$\Phi_N,\Psi_N$ contains the matrix $T_{N+1}$ which may grow (if it does not
grow, $H$ is essentially self-adjoint and the limit Wronskian indeed vanishes).

\vspace{.2cm}

The next aim is to provide a detailed description of the deficiency spaces and
certain isotropic subspaces associated to them.

\vspace{.2cm}

\begin{theo}
\label{theo-domainchar} The sum of the deficiency spaces $N_z\dotplus
N_{\overline{z}}$ can be characterized as follows:
\begin{eqnarray}
\Dd(\HH^*)\,/\,\Dd(\HH) & = & \tilde{\psi}^{\mbox{\rm\tiny
D},z}\;\CC^L\;\dotplus\;\tilde{\psi}^{\mbox{\rm\tiny D},\overline{z}}\;\CC^L
\label{eq-char1}
\\
& = & \left\{\, \phi\in\Dd(\HH^*)\; \left| \;\,\lim_{N\to\infty}\,\Ww(\Phi_N
,\Psi_N)\neq 0\;\;\mbox{\rm for some
}\psi\in\Dd(\HH^*) \;\right.\right\} \label{eq-char2} \\
& = & \psi^z\;(\Pp^z_-+\Pp^z_+)\;\CC^{2L}\;=\;\hat{\psi}^z\;\CC^{2L}\;.
\label{eq-char3}
\end{eqnarray}

\end{theo}

\vspace{.2cm}

\noindent {\bf Proof.} From the above $N_z=\tilde{\psi}^{\mbox{\rm\tiny
D},z}\CC^L$ and therefore \eqref{eq-char1} follows directly from
Proposition~\ref{prop-domain}. Instead of \eqref{eq-char2}, let us now show
$$
\Dd(\HH)\;=\; \left\{\, \phi\in\Dd(\HH^*)\; \left|
\;\,\lim_{N\to\infty}\,\Ww(\Phi_N ,\Psi_N)= 0\;\;\mbox{\rm for all
}\psi\in\Dd(\HH^*) \;\right.\right\}\;.
$$
Hence let $\phi\in\Dd(\HH)$ and $\psi\in\Dd(\HH^*)$. By
Proposition~\ref{prop-limitwronskian} the property $\Ww(\Phi_N,\Psi_N)\to 0$ as
$N\to\infty$ is equivalent to having $\langle\phi\,|\,\HH^*\psi\rangle-
\langle\HH^*\phi\,|\,\psi\rangle=0$. Now by the definition of $\Dd(\HH)$, there
exists a sequence $(\phi^{(k)})_{k\geq 1}$ of compactly supported
$\phi^{(k)}\in\Hh$ such that $\lim_{k\to\infty}\phi^{(k)}=\phi$ and that
$(H\phi^{(k)})_{k\geq 1}$ is a Cauchy sequence. Then
$\HH\phi=\lim_{k\to\infty}H\phi^{(k)}$. As $\HH^*$ is an extension of $\HH$,
one thus has
$$
\langle\phi\,|\,\HH^*\psi\rangle\,-\, \langle\HH^*\phi\,|\,\psi\rangle\;=\;
\lim_{k\to\infty}\,\bigl( \langle\phi^{(k)}\,|\,\HH^*\psi\rangle\,-\,
\langle\HH^*\phi^{(k)}\,|\,\psi\rangle\bigr)\; =\;
\lim_{k\to\infty}\,\lim_{N\to\infty}\,\Ww((\Phi^{(k)})_N,\Psi_N)\;,
$$
where the second equality follows from Proposition~\ref{prop-limitwronskian}.
But as $\phi^{(k)}$ is compactly supported, $\Ww((\Phi^{(k)})_N,\Psi_N)=0$ for
$N$ sufficiently large. This shows the inclusion $\subset$. For the converse,
suppose $\phi\in\Dd(\HH^*)$ satisfies $\Ww(\Phi_N ,\Psi_N)\to 0$ as
$N\to\infty$ for all $\psi\in\Dd(\HH^*)$. By
Proposition~\ref{prop-limitwronskian} this is equivalent to $\langle
\phi|\HH^*\psi\rangle=\langle\HH\phi|\psi\rangle$ for all $\psi\in\Dd(\HH^*)$.
Define compactly supported $\phi^{(k)}$ by $\phi^{(k)}_n=\delta_{n\leq
k}\phi_n$. Clearly $\phi^{(k)}\to\phi$ as $k\to\infty$. It remains to show that
$H\phi^{(k)}\to \HH\phi$ as $k\to\infty$; then $\phi\in\Dd(\HH)$. As
$\Dd(\HH^*)$ is dense, it is sufficient to show that
$\langle\psi|(H\phi^{(k)}-\HH\phi)\rangle\to 0$ as $k\to\infty$. But this
follows from
$\langle\psi|H\phi^{(k)}\rangle=\langle\HH^*\psi|\phi^{(k)}\rangle$ which again
results from Proposition~\ref{prop-limitwronskian}.

\vspace{.1cm}

Finally let us prove \eqref{eq-char3}, first the inclusion $\supset$. Let
$\phi=\psi^z(\Pp_+^z+\Pp^z_-)v\neq 0$ for some $v\in\CC^{2L}$. By construction,
$\phi\in\Hh$ and $H\phi=z\phi$ is thus also in $\Hh$ so that
$\phi\in\Dd(\HH^*)$. Furthermore, consider
$\psi=\psi^z(\Pp_+^z+\Pp^z_-)w\in\Dd(\HH^*)$ with $w\in\CC^{2L}$. Then
$$
\lim_{N\to\infty}\,\Ww(\Phi_N ,\Psi_N)\;=\;\lim_{N\to\infty}\,v^*
\,(\Pp_+^z+\Pp^z_-)\,\Qq^z_N(\Pp_+^z+\Pp^z_-)\,w\;.
$$
The r.h.s. does not vanish for adequate choice of $w$. Hence
$\phi\in\Dd(\HH^*)/\Dd(\HH)$. The equality now follows from
Proposition~\ref{prop-dimensions}(iv) and (v) which shows that the dimension of
$\psi^z(\Pp^z_-+\Pp^z_+)\CC^{2L}$ is equal to
$n_z+n_{\overline{z}}=\dim(\Dd(\HH^*)/\Dd(\HH))$. \hfill $\Box$

\vspace{.2cm}

\begin{coro}
\label{coro-domainchar} $\psi^z\Pp^z_0w\in\Dd(\HH)$ for all $w\in\CC^{2L}$.
Therefore $\lim_{N\to\infty}\Ww(\Psi^z_N\Pp^z_0,\Phi_N)=0$ for all
$\phi\in\Dd(\HH^*)$.
\end{coro}

\begin{proposi}
\label{prop-limitiso} Let $\Im m(z)\neq 0$.

\vspace{.1cm}

\noindent {\rm (i)} Let $\Phi$ be a $2L\times n$-matrix. Then
$$
\Phi \;\;\mbox{\rm isotropic for } \Qq^z\qquad\Longleftrightarrow \qquad
\lim_{N\to\infty}\,\Ww(\Psi^z_N\Phi ,\Psi_N^z\Phi)\,=\, \nul\;.
$$

\vspace{.1cm}

\noindent {\rm (ii)} Let $\phi=(\phi^{(1)},\ldots,\phi^{(n)})$ with
$\phi^{(k)}\in N_z\dotplus N_{\overline{z}}$ for $k=1,\ldots,n$. Then
$$
({\psi}^z {\Rr}^z)^*\phi \;\;\mbox{\rm isotropic for } \Qq^z
\qquad\Longleftrightarrow \qquad \lim_{N\to\infty}\,\Ww(\Phi_N ,\Phi_N)\,=\,
\nul \;.
$$
\end{proposi}

\vspace{.2cm}

\noindent {\bf Proof.} (i) is just a reformulation of the definition. (ii) Due
to Theorem~\ref{theo-domainchar}, $\phi=\psi^z{\Rr}^z a$ for the $2L\times
n$-matrix $a= (z-\overline{z})({\psi}^z)^*\phi$. Hence
$$
\lim_{N\to\infty}\,\Ww(\Phi_N ,\Phi_N)\;=\; ({\Rr}^z a)^*\Qq^z ({\Rr}^z a)\;=\;
|z-\overline{z}|^2\,\bigl((\psi^z\Rr^z)^*\phi\bigr)^*
\Qq^z\bigl((\psi^z\Rr^z)^*\phi\bigr)\;,
$$
which implies the result. \hfill $\Box$

\vspace{.2cm}

The following theorem shows that isotropic subspaces for $\Qq^z$ are naturally
linked to isotropic subspaces of $N_z\oplus N_{\overline{z}}$ w.r.t. to the
standard form $\Gg=\,$diag$(\one,-\one)$.

\vspace{.2cm}

\begin{theo}
\label{theo-isochar} Suppose $\Im m(z)\neq 0$. Let $\Phi$ be a $2L\times
n$-matrix. Then
\begin{equation}\label{eq-isochar}
\Phi \;\;\mbox{\rm isotropic for } \Qq^z\qquad\Longleftrightarrow \qquad
\begin{array}{c}
\psi^z\Phi\,\CC^{n}\;=\;\psi^z\Pp^z_0\Phi\,\CC^{n} +
(\tilde{\psi}^{\mbox{\rm\tiny D},\overline{z}}V-\tilde{\psi}^{\mbox{\rm\tiny
D},{z}})\;\CC^L\;\\
\mbox{\rm for a partial isometry }V=P^{\overline{z}}_+VP^z_+\;.
\end{array}
\end{equation}
If, moreover, $\Im m(z)>0$, $n_z\leq n_{\overline{z}}$ and $[\Phi]_\sim\in
\GM_L$, then $V$ in {\rm \eqref{eq-isochar}} is given by
$$ V\;=\; (\imath W)^*\;,
$$
where
$-\pi([\Phi]_\sim)=S^z+(R^z)^{\frac{1}{2}}W(-R^{\overline{z}})^{\frac{1}{2}}$.
\end{theo}

\vspace{.2cm}

\noindent {\bf Proof.} If $\Phi$ is isotropic for $\Qq^z$, then
$\Pp^z_\infty\Phi=\nul$. Hence by Theorem~\ref{theo-domainchar} one has
$\psi^z\Phi=\psi^z\Pp^z_0\Phi+\tilde{\psi}^{\mbox{\rm\tiny
D},\overline{z}}\alpha+\tilde{\psi}^{\mbox{\rm\tiny D},{z}}\beta$ for two
$L\times n$-matrices $\alpha=P^{\overline{z}}_+\alpha$ and $\beta=P^z_+\beta$.
Due to Proposition~\ref{prop-limitiso}(i) and Corollary~\ref{coro-domainchar},
the isotropy of $\Phi$ leads to
$$
\lim_{N\to\infty}\;\Ww (\tilde{\Psi}_N^{\mbox{\rm\tiny
D},\overline{z}}\alpha+\tilde{\Psi}_N^{\mbox{\rm\tiny D},{z}}\beta,
\tilde{\Psi}_N^{\mbox{\rm\tiny
D},\overline{z}}\alpha+\tilde{\Psi}_N^{\mbox{\rm\tiny D},{z}}\beta)
\;=\;\nul\;.
$$
By Proposition~\ref{prop-limitWronskianformulas} this implies
$\alpha^*P^{\overline{z}}_+\alpha=\beta^*P^z_+\beta$ or simply
$\alpha^*\alpha=\beta^*\beta$. If now $Q$ denotes the projection on the
orthogonal complement of the kernel of $\alpha^*\alpha$, then $V=-(\beta
\alpha^{-1}Q)^*$ is a partial isometry satisfying $\alpha=-V\beta$ and hence
the r.h.s. of \eqref{eq-isochar}. The inverse implication of \eqref{eq-isochar}
follows from the same calculation.

\vspace{.1cm}

Now let $\Im m(z)>0$, $n_z\leq n_{\overline{z}}$ and suppose that the isotropic
plane $\Phi$ is maximal, namely $[\Phi]_\sim\in \GM_L$. By
Proposition~\ref{prop-Weylsurfacedescription} and
Definition~\ref{def-limitsurface}, it follows that $G=-\pi([\Phi]_\sim)\in
\partial_\mx\WM^z$ and that $[\Phi]_\sim=[\Phi_G]_\sim$. Thus, by
\eqref{eq-isochar} and Theorem~\ref{theo-domainchar},
$$
-\;{\psi}^{\mbox{\rm\tiny D},{z}}G\,+\,{\psi}^{\mbox{\rm\tiny A},{z}}
\;=\;\psi^z\Pp^z_0\Phi_G\,+\,(\tilde{\psi}^{\mbox{\rm\tiny
D},\overline{z}}V\,-\,\tilde{\psi}^{\mbox{\rm\tiny D},{z}})c \;,
$$
for some $L\times L$-matrix $c$ for which the range is $P^z_+\CC^L$ and $V$ is
injective on $P^z_+\CC^L$. Now let us place the two sides as first arguments
into $\lim_{N\to\infty}\,\Ww(\,.\, , \tilde{\Psi}_N^{\mbox{\rm\tiny
D},\overline{z}})$ and $\lim_{N\to\infty}\,\Ww(\,.\, ,
\tilde{\Psi}_N^{\mbox{\rm\tiny D},{z}})$. Appealing again to
Corollary~\ref{coro-domainchar} and the identities in
Proposition~\ref{prop-limitWronskianformulas}, one finds
$$
-\,\imath\,\bigl((\overline{z}-z)R^{\overline{z}}\bigr)^{\frac{1}{2}}P^{\overline{z}}_+
\;=\;(\overline{z}-z) \,c^*\,V^*\,P^{\overline{z}}_+\;,
$$
and
$$
-\,G^*\,(z-\overline{z})
\bigl((z-\overline{z})R^{{z}}\bigr)^{-\frac{1}{2}}P^{{z}}_+ \,+\,
(z-\overline{z})S^{\overline{z}}\bigl((z-\overline{z})R^z\bigr)^{-\frac{1}{2}}P^z_+
\;=\;-\,(z-\overline{z})\,c^*\,P^z_+\;.
$$
Because $G=S^r+(R^z)^{\frac{1}{2}}W(-R^{\overline{z}})^{\frac{1}{2}}$ and
$(S^z)^*=S^{\overline{z}}$, simplifying gives
$$
\imath\,(-R^{\overline{z}})^{\frac{1}{2}}\,P^{\overline{z}}_+ \;=\;
(z-\overline{z})^{\frac{1}{2}} \,c^*\,V^*\,P^{\overline{z}}_+\;, \qquad
(-R^{\overline{z}})^{\frac{1}{2}}\,W^*\,P^{{z}}_+ \;=\;
(z-\overline{z})^{\frac{1}{2}} \,c^*\,P^{{z}}_+\;.
$$
As $V^*=P^{{z}}_+V^*$, the second equation can be replaced in the first to
complete the proof. \hfill $\Box$

\section{Self-adjointness in the limit point case} \label{sec-limitpoint}

Here we consider the case $n_z=n_{\overline{z}}=0$, namely that $H$ is in the
limit point case. Proposition~\ref{prop-Greenestimates} gives an efficient
criterion for the limit point case. Then $\HH=\HH^*$ is self-adjoint and by
strong resolvent convergence ($H^N_\xi$ converges weakly to the self-adjoint
operator $\HH$), Green's matrix of $\HH$ is  the limit point in the literal
sense given by
$$
G^z\;=\;\pi^*_1(\HH-z\,\one)^{-1}\pi_1 \;=\; \lim_{N\to\infty}\;
G_N^z(\xi)\;=\; \lim_{N\to\infty}\; S_N^z\;=\; S^z\;.
$$
In particular, this convergence is independent of the right boundary condition
$\xi$. Moreover, by Proposition~\ref{prop-deficiency} there is only a single
square-integrable matricial solution $\phi$ (that is, $\phi v\in\Hh$ for all
$v\in\CC^L$) of $(\HH-z\,\one)\phi=\pi_1$. This solution is given in equation
\eqref{eq-actDir0} with $G=G^z$. One also deduces the weak convergence of the
spectral measures which can be introduced using the Herglotz representation
theorem \cite{GT}. Associated to the Herglotz functions $z\mapsto G^z_N(\xi)$
and $z\mapsto G^z$ there are matrix-valued spectral measures $\rho_N^\xi$ and
$\rho$ by
\begin{equation}
\label{eq-specme} G^z_N(\xi)\;=\;\int\rho_N^\xi(dE)\;\frac{1}{E-z}\;, \qquad
G^z\;=\;\int\rho(dE)\;\frac{1}{E-z}\;.
\end{equation}
Then $\rho_N^\xi$ converges weakly to $\rho$. A further result on finite volume
approximation is the following proven in \cite{FHS}. The definition of the
M\"obius transformation (designated by a dot) is recalled in the appendix.

\begin{proposi}
\label{prop-Greenconvergence} Let $H$ be in the limit point case. Then, for any
sequence $Z_N\in\UM_L$, one has
$$
G^z\;=\;-\,\lim_{N\to\infty} \;\Tt^z(N,0)^{-1}\cdot (-\,Z_N)\;.
$$
\end{proposi}

\vspace{.2cm}

\noindent {\bf Proof.}. This follows directly from Theorem~\ref{theo-Weyl} and
the proof of Proposition~\ref{prop-basicformulas}. \hfill $\Box$

\vspace{.2cm}

The convergence of $S^z_N$ to $G^z$ is a particular case of this result because
$$
S_N^z\;=\; -\;\Tt^z(N,0)^{-1}\cdot (A^z_N(C^z_N)^{-1})^*\;,
$$
and $A^z_N(C^z_N)^{-1}\in\UM_L$ by Proposition~\ref{prop-Greenformulas}(iii).

\section{Maximal symmetric extensions in the limit surface case}
\label{sec-extensions}

Now we turn to the case of non-vanishing deficiency indices and suppose that
$n_z\leq n_{\overline{z}}$ throughout. A maximal symmetric extension of $\HH$
then has deficiency indices $(0,n_{\overline{z}}-n_z)$ and the extensions are
self-adjoint precisely when $n_z=n_{\overline{z}}$. The latter is always the
case {\it e.g.} for real $H$ because then $R^z=-\overline{R^{\overline{z}}}$
(von Neumann's conjugation theorem directly leads to the same conclusion). Let
us first recall von Neumann's construction of these extensions using the
deficiency spaces $N_\zeta$ and $N_{\overline{\zeta}}$ for some fixed
$\zeta\in\CC$ with $\Im m(\zeta)>0$ (often, the choice $\zeta=\imath$ is made).
The isomorphism of $N_\zeta$ with $\mbox{\rm
Ker}(R^\zeta)^\perp=P^\zeta_+\CC^L$ as given in \eqref{eq-defiso} will be used,
hence we suppose given a partial isometry
$V=P_+^{\overline{\zeta}}VP_+^{{\zeta}}:\CC^L\to\CC^L$. Thus $V^*V=P^\zeta_+$,
but the projection $VV^*$ satisfies $VV^*\leq P^{\overline{\zeta}}_+$ with
equality if and only if the deficiency indices are equal. It follows that
$\tilde{\psi}^{\mbox{\rm\tiny D},\overline{\zeta}}V
(\tilde{\psi}^{\mbox{\rm\tiny D},\zeta})^*:N_\zeta\to N_{\overline{\zeta}}$ is
also a partial isometry. As $(\HH-\overline{\zeta}\,\one)^{-1}: \mbox{\rm
Ran}(\HH-\overline{\zeta}\,\one)\to\Dd(\HH)$ and
$(\HH-{\zeta}\,\one):\Dd(\HH)\to \mbox{\rm Ran}(\HH-{\zeta}\,\one)$, one can,
using \eqref{eq-Ndef}, define the following bounded operator on $\Hh$:
\begin{equation}\label{eq-UWdef}
\UU_V\;=\;(\HH-{\zeta}\,\one)(\HH-\overline{\zeta}\,\one)^{-1}\oplus
\tilde{\psi}^{\mbox{\rm\tiny
D},\overline{\zeta}}\,V\,(\tilde{\psi}^{\mbox{\rm\tiny D},\zeta})^*\;.
\end{equation}
%

\begin{theo}
\label{theo-extensionconstruct} Let $n_\zeta\leq n_{\overline{\zeta}}$. The
operator $\UU_V$ is a partial isometry and $\UU_V-\one$ maps $\Hh$ onto
\begin{equation}
\label{eq-extdomain} \Dd(\HH_V)\;=\; \Dd(\HH)\oplus
\bigl(\tilde{\psi}^{\mbox{\rm\tiny
D},\overline{\zeta}}\,V-\tilde{\psi}^{\mbox{\rm\tiny D},\zeta}\bigr)\,\CC^L\;.
\end{equation}
This set is therefore a legitimate domain for the operator
\begin{equation}\label{eq-HWdef}
\HH_V\;=\;(\,\overline{\zeta}\,\UU_V-\zeta\,\one)(\UU_V-\one)^{-1}\;.
\end{equation}
Then $\HH_V$ is a maximal symmetric extension of $\HH$ which has no spectrum in
the upper half-plane. If $n_\zeta= n_{\overline{\zeta}}$, the extension $\HH_V$
is self-adjoint.
\end{theo}

\vspace{.2cm}

It might seem adequate to place an index $\zeta$ on $\UU_V$ and $\HH_V$, but we
refrain from doing so.

\vspace{.2cm}

\noindent {\bf Proof.} (These are the basic standard facts about von Neumann
extensions.) The fact that $\UU_V$ defined in \eqref{eq-UWdef} is a partial
isometry results from the orthogonal decompositions
$\Hh=\,$Ran$(\HH-\overline{\zeta}\,\one)\oplus
N_\zeta=\,$Ran$(\HH-{\zeta}\,\one)\oplus N_{\overline{\zeta}}$ because first of
all the Cayley transform $
(\HH-{\zeta}\,\one)(\HH-\overline{\zeta}\,\one)^{-1}: \,$
Ran$(\HH-\overline{\zeta}\,\one) \to\,$Ran$(\HH-{\zeta}\,\one)$ is a unitary
and second of all $\tilde{\psi}^{\mbox{\rm\tiny
D},\overline{\zeta}}V(\tilde{\psi}^{\mbox{\rm\tiny D},\zeta})^*:N_\zeta\to
N_{\overline{\zeta}}$ is a partial isometry by construction, satisfying
$\UU_V^*\UU_V=\one$ and $\UU_V\UU_V^*=\one- \tilde{\psi}^{\mbox{\rm\tiny
D},\overline{\zeta}}(P_+^{\overline{\zeta}}-VV^*)(\tilde{\psi}^{\mbox{\rm\tiny
D},\overline{\zeta}})^*$. Furthermore, for
$\phi\in\,$Ran$(\HH-\overline{\zeta}\,\one)$ and $v\in\CC^L$,
\begin{equation}\label{eq-intermed2}
(\UU_V-\one)(\phi+\tilde{\psi}^{\mbox{\rm\tiny D},{\zeta}}v)\;=\;
(\overline{\zeta}-\zeta)(\HH-\overline{\zeta}\,\one)^{-1}\phi+
\tilde{\psi}^{\mbox{\rm\tiny D},\overline{\zeta}}Vv -
\tilde{\psi}^{\mbox{\rm\tiny D},{\zeta}}v \;.
\end{equation}
In particular, $\UU_V-\one:\Hh\to\Dd(\HH_V)$ with $\Dd(\HH_V)$ defined by
\eqref{eq-extdomain}. Hence $\HH_V$ given in \eqref{eq-HWdef} with domain
$\Dd(\HH_V)$ is well-defined. It remains to check that $\HH_V$ is a symmetric
extension of $\HH$ for which the resolvent set contains the upper half-plane.
Setting $\psi=(\overline{\zeta}-\zeta)(\HH-\overline{\zeta}\,\one)^{-1}\phi\in
\Dd(\HH)$ in \eqref{eq-intermed2}, the action of $\HH_V$ on its domain is
explicitly given by
\begin{eqnarray*}
\HH_V\bigl(\psi+\tilde{\psi}^{\mbox{\rm\tiny
D},\overline{\zeta}}\,V\,v-\tilde{\psi}^{\mbox{\rm\tiny D},\zeta}v\bigr) & = &
(\overline{\zeta}\,\UU_V-\zeta\,\one)\,\Bigl(\,\frac{1}{\overline{\zeta}-\zeta}\,
(\HH-\overline{\zeta}\,\one) \psi+ \tilde{\psi}^{\mbox{\rm\tiny D},\zeta}v\,\Bigr) \\
& = & \HH\,\psi + \,\overline{\zeta}\,\tilde{\psi}^{\mbox{\rm\tiny
D},\overline{\zeta}}\,V\,v-\zeta\,\tilde{\psi}^{\mbox{\rm\tiny D},\zeta}v\;.
\end{eqnarray*}
Hence $\HH_V$ is an extension of $\HH$. Moreover, using this explicit formula
as well as the orthogonal decompositions Ran$(\HH-\overline{\zeta}\,\one)\oplus
N_\zeta=\,$Ran$(\HH-{\zeta}\,\one)\oplus N_{\overline{\zeta}}$, a short
calculation with all the cross terms shows the symmetry of $\HH_V$, namely
$\langle \HH_V\phi|\psi\rangle=\langle\phi|\HH_V\psi\rangle$ for all
$\phi,\psi\in\Dd(\HH_V)$. Finally, let us come to the spectrum of $\HH_V$.  For
$\lambda\neq \overline{\zeta}$, one has
$$
\HH_V-\lambda\,\one\;=\;(\overline{\zeta}-\lambda)\,
\Bigl(\,\UU_V-\frac{\zeta+\lambda}{\overline{\zeta}-\lambda}\,\Bigr)\,(\UU_V-\one)^{-1}\;.
$$
For $\Im m(\lambda)>0$ one checks that the fraction
$(\zeta+\lambda)(\overline{\zeta}-\lambda)^{-1}$ has modulus larger than $1$
(because also $\Im m(\zeta)>0$). But the spectrum of the partial isometry
$\UU_V$ lies in the closed unit disc, hence $\HH_V-\lambda\,\one$ is invertible
and $\lambda$ with $\Im m(\lambda)>0$ lies in the resolvent set of $\HH_V$. If
$n_\zeta= n_{\overline{\zeta}}$, then $\UU_V$ is unitary and its spectrum lies
on the circle $S^1$. As the above fraction never lies on $S^1$, one can show as
above that all $\lambda\in\CC/\RR$ are in the resolvent set, except for
$\lambda=\overline{\zeta}$. But in the latter case,
$\HH_V-\overline{\zeta}\,\one=(\zeta-\overline{\zeta})(\UU_V-\one)^{-1}$ is
also invertible, so also $\overline{\zeta}$ is in the resolvent set of $\HH_V$.
\hfill $\Box$

\vspace{.2cm}

If $H$ is real, then it is well-known that the extension $\HH_V$ is also
time-reversal invariant only if $V$ is symmetric (this can be checked in the
above argument). The link of the von Neumann theory of extensions to the Weyl
theory is based on the following result.

\begin{theo}
\label{theo-isospaces} Let $\zeta$, $V$ and $n_\zeta\leq n_{\overline{\zeta}}$
be as above, and $z\in\CC/\RR$. Then
$(\psi^z\Rr^z)^*(\tilde{\psi}^{\mbox{\rm\tiny
D},\overline{\zeta}}V-\tilde{\psi}^{\mbox{\rm\tiny D},\zeta})\CC^L$ is an
$n_z$-dimensional isotropic subspace for $\Qq^z$. It has a unique extension to
a maximally isotropic subspace of $\Qq^z$ which is given by
\begin{equation}\label{eq-zplane}
\Pp^z_0\,\CC^{2L}\;+\;(\psi^z\Rr^z)^*(\tilde{\psi}^{\mbox{\rm\tiny
D},\overline{\zeta}}V-\tilde{\psi}^{\mbox{\rm\tiny D},\zeta})\,\CC^L\;.
\end{equation}
If $\Im m(z)>0$, this subspace is of dimension $L$. For $\Im m(z)<0$, it is
$L$-dimensional if and only if the deficiency indices are equal. Let the plane
{\rm \eqref{eq-zplane}} be given by $\Phi^z_V\CC^L$ for some $2L\times
L$-matrix $\Phi^z_V$. Then
\begin{equation}\label{eq-domainrep}
\Dd(\HH_V)\;=\;\Dd(\HH)\,+\,\psi^z\,\Phi^z_V\,\CC^L\;.
\end{equation}
\end{theo}

\vspace{.2cm}

\noindent {\bf Proof.} First let us note that
$\phi=\tilde{\psi}^{\mbox{\rm\tiny
D},\overline{\zeta}}V-\tilde{\psi}^{\mbox{\rm\tiny D},\zeta}$ is given by
vectors in $N_z\dotplus N_{\overline{z}}=N_\zeta\dotplus N_{\overline{\zeta}}$
due to Theorem~\ref{theo-domainchar}. By Proposition~\ref{prop-limitiso} one
hence has to show that the limit Wronskian of $\phi$ vanishes. But this holds
by the same calculation as in the proof of Theorem~\ref{theo-isochar}. As
already pointed out in the proof of Proposition~\ref{prop-maxiso}, any
$n_z$-dimensional isotropic subspace for $\Qq^z$ which lies in
$(\Pp^z_++\Pp^z_-)\CC^{2L}$ is completed by $\Pp^z_0$ to a maximally isotropic
subspace of $\Qq^z$. All the claims on the dimension of the subspace therefore
directly follow from Proposition~\ref{prop-maxiso}.

\vspace{.1cm}

For the proof of \eqref{eq-domainrep}, let us use the projection
$\tilde{\psi}^z(\tilde{\psi}^z)^*=(z-\overline{z})\psi^z(\psi^z\Rr^z)^*$. It
satisfies $\tilde{\psi}^z(\tilde{\psi}^z)^*\psi^z\Pp^z_0=\psi^z\Pp^z_0$ and
$\tilde{\psi}^z(\tilde{\psi}^z)^*\phi=\phi$ for all $\phi\in N_z\dotplus
N_{\overline{z}}$. For $v\in\CC^L$, let $\phi=(\tilde{\psi}^{\mbox{\rm\tiny
D},\overline{\zeta}}V-\tilde{\psi}^{\mbox{\rm\tiny D},\zeta})v\in\Dd(\HH_V)\cap
(N_z\dotplus N_{\overline{z}})$. Then
$$
\phi\;=\;\tilde{\psi}^z(\tilde{\psi}^z)^*\phi\;=\;
\psi^z(z-\overline{z})(\psi^z\Rr^z)^*(\tilde{\psi}^{\mbox{\rm\tiny
D},\overline{\zeta}}V-\tilde{\psi}^{\mbox{\rm\tiny
D},\zeta})v\;=\;\psi^z\Phi^z_Vw\;,
$$
for some $w\in\CC^{L}$. This proves the inclusion $\subset$ of
\eqref{eq-domainrep}. For the converse, recall that one has
$\psi^z\Pp^z_0\CC^{2L}\subset \Dd(\HH)$ by Corollary~\ref{coro-domainchar}. The
second part of \eqref{eq-zplane} also leads to a contribution in $\Dd(\HH_V)$
because $\psi^z(\psi^z\Rr^z)^*$ is proportional to the projection
$\tilde{\psi}^z(\tilde{\psi}^z)^*$. \hfill $\Box$

\vspace{.2cm}

If the deficiency indices are equal, one can characterize  the domain of the
symmetric extension using only the limit Wronskian.

\vspace{.2cm}

\begin{proposi}
\label{prop-domainextension} For equal deficiency indices, the domain of
$\HH_V$ given by {\rm \eqref{eq-extdomain}} is
\begin{equation}
\label{eq-extdomain2} \Dd(\HH_V)\;=\; \left\{\;\phi\in\Dd(\HH^*)\;\left|\;
\lim_{N\to\infty}\,\Ww\bigl(\Phi_N, \tilde{\Psi}_N^{\mbox{\rm\tiny
D},\overline{\zeta}}\,V-\tilde{\Psi}_N^{\mbox{\rm\tiny
D},\zeta}\bigr)=\nul\;\right.\right\}\;.
\end{equation}
\end{proposi}

\vspace{.2cm}

\noindent {\bf Proof.} Let $\tilde{\Dd}$ be the set on the r.h.s. of
\eqref{eq-extdomain2}. Because of Theorem~\ref{theo-domainchar} one has
$$
\tilde{\Dd} \;=\;\Dd(\HH)\;\dotplus\;\left\{\;\phi\in N_\zeta\dotplus
N_{\overline{\zeta}} \;\left|\; \lim_{N\to\infty}\,\Ww\bigl(\Phi_N,
\tilde{\Psi}_N^{\mbox{\rm\tiny
D},\overline{\zeta}}\,V-\tilde{\Psi}_N^{\mbox{\rm\tiny
D},\zeta}\bigr)=\nul\;\right.\right\}\;.
$$
This has to be compared with $\Dd(\HH_V) =\Dd(\HH)\oplus
\bigl(\tilde{\psi}^{\mbox{\rm\tiny
D},\overline{\zeta}}V-\tilde{\psi}^{\mbox{\rm\tiny D},\zeta}\bigr)\CC^L$ as
defined in {\rm \eqref{eq-extdomain}}. By the proof of
Theorem~\ref{theo-isospaces}, $\Dd(\HH_V)\subset \tilde{\Dd}$. On the other
hand, $N_\zeta\dotplus N_{\overline{\zeta}}=\tilde{\psi}^{\mbox{\rm\tiny
D},\zeta}\CC^L+\tilde{\psi}^{\mbox{\rm\tiny D},\overline{\zeta}}\CC^L$. Hence
let $\tilde{\psi}^{\mbox{\rm\tiny D},\zeta}v+\tilde{\psi}^{\mbox{\rm\tiny
D},\overline{\zeta}}w\in\tilde{\Dd}$ for $v,w\in\CC^L$ such that $P^\zeta_+v=v$
and $P^{\overline{\zeta}}_+w=w$, namely
$$
\lim_{N\to\infty}\,\Ww\bigl(\tilde{\Psi}_N^{\mbox{\rm\tiny
D},\zeta}v+\tilde{\Psi}_N^{\mbox{\rm\tiny D},\overline{\zeta}}w,
\tilde{\Psi}_N^{\mbox{\rm\tiny
D},\overline{\zeta}}\,V-\tilde{\Psi}_N^{\mbox{\rm\tiny D},\zeta}\bigr)=\nul\;.
$$
Proceeding as in Theorem~\ref{theo-isospaces}, this implies
$v^*P^\zeta_++w^*P^{\overline{\zeta}}_+V=\nul$, that is $Vv=-VV^*w=-w$, the
latter because the deficiency indices are equal and therefore
$VV^*=P^{\overline{\zeta}}_+$. \hfill $\Box$

\vspace{.2cm}

\section{The resolvent of the extensions}
\label{sec-extensionresolvent}

Using the prior results it is possible to calculate the Green function of the
maximal symmetric extension $\HH_V$ constructed in the last section. It is
defined by
$$
G^z_V\;=\; \pi_1^*(\HH_V-z\,\one)^{-1}\pi_1\;, \qquad \Im m(z)>0\;.
$$

\begin{theo}
\label{theo-greencalc} For $\Im m(z)>0$ and $n_z\leq n_{\overline{z}}$, one has
\begin{equation}
\label{eq-greencalc} G^z_V \;=\;
S^z\,+\,(R^z)^{\frac{1}{2}}W^z_V(-R^{\overline{z}})^{\frac{1}{2}}\;,
\end{equation}
where $W^z_V$ is a partial isometry from $\mbox{\rm
Ker}(R^{\overline{z}})^\perp$ to $\mbox{\rm Ker}(R^{z})^\perp$ given explicitly
below. One has
\begin{equation}
\label{eq-greencalc2} W^\zeta_V\;=\;-\;\imath\; V^*\;.
\end{equation}

\end{theo}

\vspace{.2cm}

\noindent {\bf Proof.} For the calculation of Green's matrix $G^z_V$ one has to
solve the equation $(\HH_V-z\,\one)\phi=\pi_1$ for $L\times L$ matricial
solutions $\phi$ such that $\phi v\in\Dd(\HH_V)$ for all $v\in\CC^L$ and $\phi$
is of maximal rank $L$. In case this solution is unique, one has then
$G^z_V=\phi_1$. By Proposition~\ref{prop-solutions}, any such solution $\phi$
is of the form $\phi= {\psi}^{\mbox{\rm\tiny D},{z}}G- {\psi}^{\mbox{\rm\tiny
A},{z}}=-\,\psi^z \Phi_G$ and then $G^z_V=-\,\pi([\Phi_G]_\sim)$. However,
there is only a unique adequate $G$ such that $\psi^z \Phi_Gv\in\Dd(\HH_V)$ for
all $v\in\CC^L$. Indeed, by Theorem~\ref{theo-isospaces}, the only
$L$-dimensional plane $[\Phi]_\sim\in\GM_L$ such that $\psi^z\Phi
v\in\Dd(\HH_V)$ for all $v\in\CC^L$ is the plane $\Phi^z_V$ given in
\eqref{eq-zplane}. Moreover, this plane is isotropic for $\Qq^z$. But by
Proposition~\ref{prop-Weylsurfacedescription} and
Definition~\ref{def-limitsurface}, $[\Phi^z_V]_\sim=[\Phi_G]_\sim$ for a unique
$G\in \UM_L$. As $\pi([\Phi_G]_\sim)=\pi([\Phi^z_V]_\sim)$, it follows that
Green's matrix is given by
$$
G^z_V\;=\;-\,\pi([\Phi^z_V]_\sim)\;.
$$
In particular, $G^z_V\in \partial_\mx\WM^z$. Now by \eqref{eq-Weyllimitrep},
there is indeed a partial isometry $W^z_V:\mbox{\rm
Ker}(R^{\overline{z}})^\perp\to\mbox{\rm Ker}(R^{z})^\perp$ such that
\eqref{eq-greencalc} holds. It clearly satisfies \eqref{eq-greencalc2} due to
Theorem~\ref{theo-domainchar}. \hfill $\Box$

\vspace{.2cm}

We now appeal to Theorems~\ref{theo-isochar} and \ref{theo-isospaces} in order
to calculate $W^z_V$ more explicitly. For this purpose, let us consider $\phi=
\tilde{\psi}^{\mbox{\rm\tiny D},\overline{\zeta}}V-\tilde{\psi}^{\mbox{\rm\tiny
D},\zeta}$ as the generator of an $n_\zeta$-dimensional subspace $\phi\,\CC^L$
of $N_\zeta\dotplus N_{\overline{\zeta}}$. The aim is to calculate $W^z_V$ such
that $\phi\,\CC^L= (\tilde{\psi}^{\mbox{\rm\tiny D},\overline{z}}(\imath
W^z_V)^* -\tilde{\psi}^{\mbox{\rm\tiny D},{z}})\CC^L$. This will be done using
an $n_{\overline{\zeta}}$-dimensional subspace $\phi_\perp\CC^L$ in
$N_\zeta\dotplus N_{\overline{\zeta}}$ which is orthogonal to $\phi\,\CC^L$
w.r.t. to the scalar product in $\Hh$. Because of
Theorem~\ref{theo-domainchar}, it is of the form
\begin{equation}\label{eq-ortho}
\phi_\perp\;=\;\tilde{\psi}^{\mbox{\rm\tiny
D},\overline{\zeta}}\alpha\,+\,\tilde{\psi}^{\mbox{\rm\tiny D},\zeta}\beta
\,+\, \left[\,\tilde{\psi}^{ \mbox{\rm\tiny
D},\overline{\zeta}}(P^{\overline{\zeta}}_+-VV^*)\,-\,
\tilde{\psi}^{\mbox{\rm\tiny D},{\zeta}}\,\gamma\,\right]\;,
\end{equation}
with $L\times L$-matrices $\alpha=VV^*\alpha$ and
$\beta=P^\zeta_+\beta=V^*V\beta$, and $\gamma=P^\zeta_+\gamma$ is of rank
$n_{\overline{\zeta}}-n_\zeta$ chosen such that the last term in brackets is
orthogonal to $\phi$ (for equal deficiency indices $\gamma=\nul$). Let us first
calculate $\gamma$. The orthogonality means
$$
\nul\;=\;\phi^*\left[\, \tilde{\psi}^{\mbox{\rm\tiny
D},\overline{\zeta}}(P^{\overline{\zeta}}_+-VV^*)\,-\,
\tilde{\psi}^{\mbox{\rm\tiny D},{\zeta}}\,\gamma\,\right]\;=\;-\,(
\tilde{\psi}^{\mbox{\rm\tiny D},{\zeta}})^* \tilde{\psi}^{\mbox{\rm\tiny
D},\overline{\zeta}}(P^{\overline{\zeta}}_+-VV^*)\,-\, V^*(
(\tilde{\psi}^{\mbox{\rm\tiny D},\overline{\zeta}})^*
\tilde{\psi}^{\mbox{\rm\tiny D},{\zeta}}-V)\,\gamma\;.
$$
Now $\tilde{\psi}^{\mbox{\rm\tiny D},\overline{\zeta}}$ and
$\tilde{\psi}^{\mbox{\rm\tiny D},{\zeta}}$ are partial isometries on linearly
independent subspaces. Therefore $(\tilde{\psi}^{\mbox{\rm\tiny
D},\overline{\zeta}})^* \tilde{\psi}^{\mbox{\rm\tiny D},{\zeta}}$ has norm less
than $1$ and it follows that $V^*( (\tilde{\psi}^{\mbox{\rm\tiny
D},\overline{\zeta}})^* \tilde{\psi}^{\mbox{\rm\tiny D},{\zeta}}-V)$ is an
invertible operator on $P^{\zeta}_+\CC^L$. Thus
$$
\gamma\;=\; (V- (\tilde{\psi}^{\mbox{\rm\tiny D},\overline{\zeta}})^*
\tilde{\psi}^{\mbox{\rm\tiny D},{\zeta}})^{-1}\,V\, (
\tilde{\psi}^{\mbox{\rm\tiny D},{\zeta}})^* \tilde{\psi}^{\mbox{\rm\tiny
D},\overline{\zeta}}(P^{\overline{\zeta}}_+-VV^*)\;.
$$
In order to determine $\alpha$ and $\beta$, let us consider
$$
\nul\;=\;\phi^*\phi_\perp\;=\;(V^*- (\tilde{\psi}^{\mbox{\rm\tiny
D},{\zeta}})^*\tilde{\psi}^{\mbox{\rm\tiny D},\overline{\zeta}})\alpha\,+\,
V^*( (\tilde{\psi}^{\mbox{\rm\tiny D},\overline{\zeta}})^*
\tilde{\psi}^{\mbox{\rm\tiny D},{\zeta}}-V)\beta\;.
$$
Now for the same reason as above, the two matrices in the the parenthesis are
invertible. Therefore one may choose:
$$
\alpha\;=\;\bigl(V^*- (\tilde{\psi}^{\mbox{\rm\tiny
D},{\zeta}})^*\tilde{\psi}^{\mbox{\rm\tiny D},\overline{\zeta}}\bigr)^{-1}
\;,\qquad \beta\;=\;\bigl(V- (\tilde{\psi}^{\mbox{\rm\tiny
D},\overline{\zeta}})^*\tilde{\psi}^{\mbox{\rm\tiny D},{\zeta}}\bigr)^{-1}V\;.
$$
It remains to determine $W^z_V$ by using the orthogonality of
$(\tilde{\psi}^{\mbox{\rm\tiny D},\overline{z}}(\imath
W^z_V)^*-\tilde{\psi}^{\mbox{\rm\tiny D},z})$ to $\phi_\perp$ in $N_z\dotplus
N_{\overline{z}}=N_\zeta\dotplus N_{\overline{\zeta}}$ ({\it cf.}
Theorem~\ref{theo-isochar}). From
$\nul=\phi_\perp^*(\tilde{\psi}^{\mbox{\rm\tiny D},\overline{z}}(\imath
W^z_V)^*-\tilde{\psi}^{\mbox{\rm\tiny D},z})$ follows
$$
(\imath W^z_V)^*\;=\;\left(\phi_\perp^*\tilde{\psi}^{\mbox{\rm\tiny
D},\overline{z}}\right)^{-1}\, \phi_\perp^*\tilde{\psi}^{\mbox{\rm\tiny D},{z}}
\;,
$$
which, given \eqref{eq-ortho} and the above choices of $\alpha$, $\beta$ and
$\gamma$, is the desired explicit expression.

\vspace{.2cm}

%
%
%
%

\section*{Appendix: reminder on M\"obius transformations}
\label{sec-Moeb}

This appendix resembles the basic properties of the M\"obius transformation as
they are used in the main text. A lot of references to the literature can be
found in \cite{DPS}. The M\"obius transformation (also called canonical
transformation or fractional transformation) is defined by
\begin{equation}
\label{eq-moebius}
\Tt\cdot Z
\;=\;
(AZ+B)\,(CZ+D)^{-1}
\,,
\qquad
\Tt
\,=\,
\left(
\begin{array}{cc} A & B \\  C & D \end{array}
\right)
\;\in\;\mbox{GL}(2L,\CC)
\,,
\;\;Z\in\mbox{Mat}(L\times L,\CC)\,,
\end{equation}
whenever the appearing inverse exists. If $\Tt\in\,$SP$(2L,\CC)$ and
$Z\in\UM_L$, then $\Tt\cdot Z$ exists and is in $\UM_L$. For $\Tt$ as in
\eqref{eq-moebius} and as long as the appearing inverse exists, the inverse
M\"obius transformation is defined by
\begin{equation}
\label{eq-moebiusinv} W:\Tt \;=\; (WC-A)^{-1}\,(B-WD) \,, \qquad
W\in\mbox{Mat}(L\times L,\CC)\,.
\end{equation}
The M\"obius transformation is a left action, namely $(\Tt\Tt')\cdot
Z=\Tt\cdot(\Tt'\cdot Z)$ as long as all objects are well-defined. The inverse
M\"obius transformation is a right action in the sense of the following
proposition, the algebraic proof of which is left to the reader.

\begin{proposi}
\label{prop-Moebinv} Under the condition that all the M\"obius und inverse
M\"obius transformations as well as matrix inverses below exist, one has the
following properties.

\vspace{.2cm}

\noindent {\rm (i)} $\;\;W=\Tt\cdot Z\;\;\Leftrightarrow\;\; W:\Tt=Z$

\vspace{.1cm}

\noindent {\rm (ii)} $\;W:(\Tt\Tt')=(W:\Tt):\Tt'$

\vspace{.1cm}

\noindent {\rm (iii)} $W:\Tt=\Tt^{-1}\cdot W$

\end{proposi}

\vspace{.2cm}


\end{document}